\documentclass[letter,notitlepage]{article}
\usepackage{amssymb,amsthm} 
\usepackage{graphicx}
\usepackage{amsmath} 
\usepackage{natbib} 
\usepackage{booktabs}
\usepackage{lscape}
\usepackage{latexsym,indentfirst,ascmac}
\usepackage{comment}%
\usepackage{color}%
\usepackage{vruler}
\usepackage[top=30truemm,bottom=30truemm,left=25truemm,right=25truemm]{geometry}

\usepackage{bm}

\newtheorem{Lemma*}{Lemma A.}
\newtheorem{Definition}{Definition}

\theoremstyle{remark}
\newtheorem{Remark}{Remark}
\newtheorem{Example}{Example}

\usepackage{threeparttable} 

\usepackage{algpseudocode}
\usepackage{algorithm,algcompatible} 
\newcommand{\algrule}[1][.2pt]{\par\vskip.5\baselineskip\hrule height #1\par\vskip.5\baselineskip}

\algnewcommand\algorithmicinput{\textbf{Input:}}
\algnewcommand\INPUT{\item[\algorithmicinput]}

\algnewcommand\algorithmicoutput{\textbf{Output:}}
\algnewcommand\OUTPUT{\item[\algorithmicoutput]}

\usepackage{rotating}
\usepackage{amsmath}
\usepackage{mathtools}
\usepackage{amssymb}
\usepackage{amsthm}
\usepackage{bm}
\usepackage{bbm}
\usepackage{graphicx}
\usepackage{tikz}
\usetikzlibrary{positioning}
\usepackage{listings}
\usepackage{comment}

\newcommand{\ou}[3]{%
  \mathrel{%
    \vcenter{\offinterlineskip
      \ialign{##\cr$#1$\cr\noalign{\kern-#3}$#2$\cr}%
    }%
  }%
}
\newcommand*{\omu}[3]{\underset{#3}{\overset{#1}{#2}}}

\newcommand*{\T}{^{\top}}

\newcommand*{\isim}{\omu{\text{\tiny{ind.}}}{\sim}{}}

\newcommand*{\deq}{\omu{\text{\tiny{d}}}{=}{}}

\newcommand*{\IN}{\mathbb{N}}

\newcommand*{\IR}{\mathbb{R}}

\newcommand*{\Par}{\operatorname{Par}}

\newcommand*{\Unif}{\operatorname{U}}

\newcommand*{\N}{\operatorname{N}}

\newcommand*{\rd}{\mathrm{d}}

\newcommand*{\Prob}{\mathbb{P}}
\newcommand*{\E}{\mathbb{E}}

\newcommand*{\diag}{\operatorname{diag}}

\newcommand*{\Var}{\operatorname{Var}}
\newcommand*{\Cov}{\operatorname{Cov}}

\newcommand*{\VaR}{\operatorname{VaR}}
\newcommand*{\RVaR}{\operatorname{RVaR}}
\newcommand*{\ES}{\operatorname{ES}}
\newcommand*{\CoVaR}{\operatorname{CoVaR}}

\newcommand*{\CoES}{\operatorname{CoES}}

\newcommand{\bc}{\bm{c}}

\newcommand{\bh}{\bm{h}}

\newcommand{\bx}{\bm{x}}
\newcommand{\bX}{\bm{X}}
\newcommand{\by}{\bm{y}}
\newcommand{\bY}{\bm{Y}}

\newcommand{\bZ}{\bm{Z}}

\newcommand{\bu}{\bm{u}}
\newcommand{\bU}{\bm{U}}
\newcommand{\be}{\bm{e}}
\newcommand{\bbi}{\bm{i}}
\newcommand{\bp}{\bm{p}}

\newcommand{\bv}{\bm{v}}

\newcommand{\bzero}{\bm{0}}

\newcommand{\X}{\bm{X}}
\newcommand{\Y}{\bm{Y}}

\newcommand{\bF}{\bm{F}}

\newcommand{\sqc}[2]{#1_{1},\dots,#1_{#2}}

\newcommand{\tra}[1]{#1^{\top}}
\newcommand{\supp}[1]{\text{supp}(#1)}

\newcommand{\bmu}{\bm{\mu}}

\newcommand{\bone}{\bm{1}}

\title{Markov Chain Monte Carlo Methods for Estimating Systemic Risk Allocations}

\author{Takaaki Koike 
\footnote{Department of Statistics and Actuarial Science, University of Waterloo, Waterloo, ON, Canada, E-mail: tkoike@uwaterloo.ca},\ 
Marius Hofert
\footnote{Department of Statistics and Actuarial Science, University of Waterloo, Waterloo, ON, Canada, E-mail: marius.hofert@uwaterloo.ca}
}

\begin{document}

\maketitle
\begin{abstract}
We propose a novel framework of estimating systemic risk measures and
  risk allocations based on Markov chain Monte Carlo (MCMC) methods.  We
  consider a class of allocations whose $j$th component can be written as some
  risk measure of the $j$th conditional marginal loss distribution given the
  so-called crisis event.  By considering a crisis event as an intersection of
  linear constraints, this class of allocations covers, for example, conditional
  Value-at-Risk (CoVaR), conditional expected shortfall (CoES), VaR
  contributions, and range VaR (RVaR) contributions as special cases.  For this
  class of allocations, analytical calculations are rarely available, and
  numerical computations based on Monte Carlo (MC) methods often provide inefficient
  estimates due to the rare-event character of the crisis events.  We propose an
  MCMC estimator constructed from a sample path of a Markov chain whose
  stationary distribution is the conditional distribution given the crisis
  event.  Efficient constructions of Markov chains, such as Hamiltonian Monte
  Carlo and Gibbs sampler, are suggested and studied depending on the crisis
  event and the underlying loss distribution. The efficiency of the MCMC estimators
  is demonstrated in a series of numerical experiments.
\end{abstract}

\section{Introduction}\label{sec:introduction}
 
In portfolio risk management, \emph{risk allocation} is an essential step to
quantify the risk of each unit of a portfolio by decomposing the total risk of
the whole portfolio.  One of the most prevalent rules to determine risk
allocations is the \emph{Euler princple}, proposed by \citet{tasche1999risk} and
justified from various viewpoints such as RORAC compatibility
(\citet{tasche1999risk} and \citet{tasche2008capital}) and cooperative game
theory (\citet{denault2001coherent}).  For
the popular risk measures VaR, RVaR, and ES, Euler allocations take the form of conditional expectations of
the underlying loss random vector given a certain rare event on the total loss
of the portfolio; see \citet{tasche2001conditional} for derivations.  We call
this rare event the \emph{crisis
  event}.

The decomposition of risks is also required in the context of systemic risk
measurement.  \emph{Systemic risk} is the risk of financial distress of an
entire economy as a result of the failure of individual components of the
financial system.  To quantify such risks, various \emph{systemic risk measures}
have been proposed in the literature, such as \emph{conditional VaR (CoVaR)}
(\citet{tobias2016covar}), \emph{conditional expected shortfall (CoES)}
(\citet{mainik2014dependence}) and \emph{marginal expected shortfall (MES)}
(\citet{acharya2017measuring}).  These three measures quantify the risk of
individuals by taking the VaR, ES and expectation of the individual loss,
respectively, under some stressed scenario, that is, given the crisis event.
\citet{chen2013axiomatic}, \citet{hoffmann2016risk} and \citet{kromer2016systemic} proposed an axiomatic characterization of systemic risk
measures, where the risk of the aggregated loss in a financial system is first
measured and then decomposed into the individual economic entities.  Due to the
similarity of risk allocations with the derivation of systemic risk measures, we
refer to both of them as \emph{systemic risk allocations}.  In fact, MES coincides with
the Euler allocation of ES, and other Euler allocations can be
regarded as special cases of systemic risk measures considered in
\citet{gourieroux2013allocating}.

Calculating systemic risk allocations given an unconditional joint loss
distribution is in general challenging since analytical calculations often
require to know the joint distribution of the marginal loss and the aggregated
loss. Furthermore, MC estimation suffers from the rare-event
character of the crisis event.  For computing CoVaR, CoES and MES,
\citet{mainik2014dependence}, \citet{bernardi2017covar} and
\citet{jaworski2017conditional} derived formulas based on the copula of the
marginal and the aggregated loss; \citet{asimit2018systemic} derived asymptotic
formulas based on extreme value theory; and \citet{girardi2013systemic} estimated
CoVaR under a multivariate GARCH model.  
\citet{vernic2006multivariate}, \citet{chiragiev2007multivariate}, 
\citet{dhaene2008some} and \citet{furman2008economic}
calculated Euler allocations for specific joint
distributions.  \citet{asimit2011asymptotics} derived asymptotic formulas for
risk allocations.  \citet{furman2009weighted} and \citet{furman2018weighted}
calculated weighted allocations, which include Euler allocations as special
cases, under a Stein-type assumption.  Concerning the numerical computation of
Euler allocations, \citet{glasserman2005measuring},
\citet{glasserman2005importance} and \citet{kalkbrener2004sensible}
  considered importance sampling methods, and \citet{siller2013measuring}
proposed the Fourier transform Monte Carlo method, all specifically for credit
portfolios.  For general copula-based dependence models, analytical calculations
of systemic risk allocations are rarely available, and an estimation method is,
to the best of our knowledge, only addressed in \citet{targino2015sequential},
where sequential Monte Carlo (SMC) samplers are applied.

We address the problem of estimating systemic risk allocations under general
copula-based dependent risks in the case where the copula between the marginal
losses and the aggregated loss are not necessarily available. We consider a general
class of systemic risk allocations in the form of risk measures of a conditional
loss distribution given a crisis event, which includes CoVaR, CoES, MES and
Euler allocations as special cases.  In our proposed method, the conditional
loss distribution, called the \emph{target distribution} $\pi$, is simulated by
a Markov chain
whose stationary distribution is the desired distribution $\pi$ by sequentially
updating the sample path based on the available information from $\pi$.  While this MCMC method resembles the SMC in
\citet{targino2015sequential}, the latter requires a more complicated
implementation involving the choice of forward and backward kernels, resampling
and move steps, and even MCMC in the move steps.  Our
suggested approach directly constructs a single sophisticated Markov chain
depending on the target distribution of interest.  Applications of MCMC to
estimating risk allocations have been studied in \citet{koike2019estimation},
specifically for VaR contributions.  Our paper explores and demonstrates the
applicability of MCMC methods to a more general class of systemic risk
allocations.

Almost all MCMC methods used in practice are of the \emph{Metropolis-Hastings
  (MH)} type (\citet{metropolis1953equation} and \citet{hastings1970monte}),
where the so-called \emph{proposal distribution} $q$ generates a candidate of
the next state based on the current state. This candidate is then accepted or
rejected according to the so-called \emph{acceptance probability} to adjust the
stationary distribution to be the target distribution $\pi$.  As explained in
Section~\ref{subsec:brief:review:MCMC} below, the resulting Markov chain has
serial correlation, which adversarially affects the efficiency of the
estimator.  An efficient MCMC of MH type is such that the proposal distribution
generates a candidate which exhibits low correlation with the current state with
sufficiently large acceptance probability.  The main difficulty in constructing
such an efficient MCMC estimator for systemic risk allocations is that the
support of the target distribution $\pi$ is subject to constraints determined by
the crisis event.  For such target distributions, simple MCMC methods such as
random walk MH are not efficient since a candidate is immediately rejected if it
violates the constraints; see
Section~\ref{subsec:mcmc:formulation:estimating:systemic:allocations} for
details.

To tackle this problem, we consider two specific MCMC methods, \emph{Hamiltonian
  Monte Carlo (HMC)} (\citet{duane1987hybrid}) and the \emph{Gibbs sampler (GS)}
(\citet{geman1984stochastic} and \citet{gelfand1990sampling}).  In the HMC
method, a candidate is generated according to the so-called Hamiltonian
dynamics, which leads to a high acceptance probability and low correlation with
the current state by accurately simulating the dynamics of sufficiently long
length; see \citet{neal2011mcmc} and \citet{betancourt2017conceptual} for an
introduction to HMC. Moreover, the HMC candidates always belong to the crisis
event by reflecting the dynamics when the chain hits the boundary of the
constraints; see \citet{rujan1997playing}, \citet{pakman2014exact},
\citet{afshar2015reflection}, \citet{yi2017roll} and
\citet{chevallier2018hamiltonian} for this reflection property of the HMC
method. An alternative method to handle the constraints is the GS, in
which the chain is updated in each component.  Since all the components except
the updated one remain fixed, a componentwise update is typically subject to
weaker constraints.  As long as such componentwise updates are feasible, the GS
candidates belong to the crisis event, and the acceptance probability is always
$1$; see \citet{geweke1991efficient}, \citet{gelfand1992bayesian} and
\citet{rodriguez2004efficient} for the application of the GS to constrained
target distributions, and see \citet{gudmundsson2014markov} and
\citet{targino2015sequential} for applications to estimating risk contributions.

Our findings include efficient MCMC estimators of systemic risk allocations
achieved via HMC with reflection and GSs.  We assume that the
unconditional joint loss density is known, possibly through its marginal
densities and copula density.  Depending on the supports of the marginal loss
distributions and the crisis event, different MCMC methods are applicable.  We
find that if the marginal loss distributions are one-sided, that is, the
supports are bounded from the left, then the crisis event is typically a bounded
set and HMC shows good performance.  On the other hand, if the marginal losses
are two-sided, that is, they have both right and left tails, the crisis event is
often unbounded and the GSs perform better, provided that
random number generators of the conditional copulas are available.  Based on the
samples generated by the MC method, we propose heuristics to determine
the parameters of the HMC and GS methods, for which no manual
  interaction is required.  Since, in the MCMC method, the conditional loss
distribution of interest is directly simulated in contrast to MC where
  rejection is applied based on the unconditional loss distribution, the MCMC
method in general outperforms the MC method in terms of the sample size and thus
the standard error. This advantage of MCMC becomes more pronounced as
the probability of the crisis event becomes smaller. We demonstrate this
efficiency of the MCMC estimators of systemic risk allocations by a series of
numerical experiments.

This paper is organized as follows. The general framework of the
estimation problem of systemic risk allocations is introduced in
Section~\ref{sec:systemic:risk:allocations:estimation}.  Our class of systemic
risk allocations is proposed in Section~\ref{subsec:class:systemic:allocations}
and their estimation via the MC method is presented in
Section~\ref{subsec:MC:estimation:allocation}.
Section~\ref{sec:mcmc:estimation:systemic:risk:allocation} is devoted to MCMC
methods for estimating systemic risk allocations.  After a brief review of MCMC
methods in Section~\ref{subsec:brief:review:MCMC}, we formulate our problem of
estimating systemic risk allocations in terms of MCMC in
Section~\ref{subsec:mcmc:formulation:estimating:systemic:allocations}.  HMC and
GS for constrained target distributions are then investigated in
Sections~\ref{subsec:HMC} and~\ref{subsec:gibbs:samplers},
respectively. In Section~\ref{sec:numerical:experiments} numerical
  experiments are conducted including simulation and empirical studies, and a
  detailed comparison of MC and our introduced MCMC methods.
  Section~\ref{sec:conclusion} concludes with practical guidance and limitations
  of the presented MCMC methods.

\section{Systemic Risk Allocations and Their Estimation}\label{sec:systemic:risk:allocations:estimation}
In this section, we define a broad class of systemic risk allocations including
Euler allocations, CoVaR and CoES as special cases.  Then the MC
method is described to estimate systemic risk allocations.

\subsection{A Class of Systemic Risk Allocations}\label{subsec:class:systemic:allocations}
Let $(\Omega,\mathcal F,\Prob)$ be an atomless probability space and let
$\sqc{X}{d}$, $d\geq 2$ be random variables on this space.  The random vector
$\X=(\sqc{X}{d})$ can be interpreted as losses of a portfolio of size $d$, or
losses of $d$ economic entities in an economy over a fixed time period.
Throughout the paper, a positive value of a loss random variable represents a
financial loss, and a negative loss is interpreted as a profit.  Let $F_{\bX}$
denote the joint cumulative distribution function (cdf) of $\bX$ with marginal
distributions $\sqc{F}{d}$.  Assume that $F_{\bX}$ admits a probability density
function (pdf) $f_{\bX}$ with marginal densities $\sqc{f}{d}$.  Sklar's theorem
(\citet{nelsen2006introduction}) allows one to write
\begin{align}
  F_{\bX}(\bx)=C(F_{1}(x_{1}),\dots,F_{d}(x_{d})),\quad
  \bx=(\sqc{x}{d})\in \IR^d,\label{eq:sklar:thm}
\end{align}
where $C: [0,1]^d \rightarrow [0,1]$ is a \emph{copula} of $\bX$.  Assuming the
density $c$ of the copula $C$ to exist, $f_{\bX}$ can be written as
\begin{align*}
  f_{\bX}(\bx)=c(F_{1}(x_{1}),\dots,F_{d}(x_{d})) f_{1}(x_{1})\cdots f_{d}(x_{d}),\quad \bx \in \IR^{d}.
\end{align*}

An \emph{allocation} $ A = (\sqc{A}{d})$ is a map from a random vector $\bX$ to
$(A_1(\bX),\dots,A_d(\bX)) \in \IR^d$.  The sum $\sum_{j=1}^d A_j(\bX)$ can be
understood as the capital required to cover the total loss of the portfolio or
the economy. 
The $j$th component $A_j(\bX)$, $j=1,\dots,d$ is then the contribution of the
$j$th loss to the total capital $\sum_{j=1}^d A_j(\bX)$.  In this paper, we
consider the following class of allocations
\begin{align*}
  A^{\sqc{\varrho}{d},\mathcal C} = (A_1^{\varrho_1,\mathcal C},\dots,A_d^{\varrho_d,\mathcal C}),\quad
  A_j^{\varrho_j,\mathcal C}(\bX) = \varrho_j(X_j \ |\ \bX \in \mathcal C )
\end{align*}
where $\varrho_j$ is a map from a random variable to $\IR$ called the $j$th
\emph{marginal risk measure} for $j=1,\dots,d$, and $\mathcal C\subseteq \IR^d$
is a set called the \emph{crisis event}.  The conditioning set
$\{\bX \in \mathcal C\}$ is simply written as $\mathcal C$ if there is no
confusion.  As we now explain, this class of allocations covers well-known
allocations as special cases.  For a random variable $X\sim F$, define the
\emph{Value-at-Risk (VaR)} of $X$ at confidence level $\alpha \in (0,1]$ by
\begin{align*}
  \VaR_{\alpha}(X):=\inf\{x \in \IR: F(x) \ge \alpha\}.
\end{align*}
\emph{Range Value-at-Risk (RVaR)} at confidence levels
$0 < \alpha_1 < \alpha_2 \leq 1$ is defined by
\begin{align*}
  \RVaR_{\alpha_1,\alpha_2}(X)=\frac{1}{\alpha_2-\alpha_1}\int_{\alpha_1}^{\alpha_2}\mbox{VaR}_{\gamma}(X)\,\rd\gamma,
\end{align*}
and, if it exists, \emph{expected shortfall (ES)} at confidence level
$\alpha\in (0,1)$ is defined by $\ES_{\alpha}(X) = \RVaR_{\alpha,1}(X)$.
Note that ES is also known as C(onditional)VaR, T(ail)VaR, A(verage)VaR
  and C(onditional)T(ail)E(expectation). These risk measures are law-invariant
in the sense that they depend only on the distribution of $X$.
Therefore, we sometimes write $\varrho(F)$ instead of $\varrho(X)$.

We now define various crisis events and marginal risk measures. A typical form
of the crisis event is an intersection of a set of linear constraints
\begin{align}\label{eq:crisis:event:intersections}
  \mathcal C = \bigcap_{m=1}^M \left\{ \bh_m\T \bx \geq v_m \right\},\quad \bh_m \in \IR^d,\quad v_m \in \IR,\quad m=1,\dots, M,\quad M\in \IN.
\end{align}
Several important special cases of the crisis event of
Form~\eqref{eq:crisis:event:intersections} are provided in the following.
\begin{Definition}[VaR, RVaR and ES crisis events]
  For $S = \sum_{j=1}^d X_j$, the \emph{VaR}, \emph{RVaR} and \emph{ES crisis events} are defined by
  \begin{align*}
    \mathcal C_{\alpha}^{\operatorname{VaR}} &= \{\bx \in \IR^d\ | \ \bone_d\T \bx = \VaR_{\alpha}(S) \},\quad \alpha \in (0,1),\\
    \mathcal C_{\alpha_1,\alpha_2}^{\RVaR} &= \{\bx \in \IR^d\ | \ \VaR_{\alpha_1}(S)\leq \bone_d\T \bx \leq \VaR_{\alpha_2}(S) \},\quad 0 < \alpha_1 < \alpha_2 \leq 1,\\
    \mathcal C_{\alpha}^{\ES} &= \{\bx \in \IR^d\ | \ \VaR_{\alpha}(S)\leq \bone_d\T \bx  \}, \quad 0<\alpha <1,\quad \alpha\in(0,1),
  \end{align*}
respectively, where $\bone_d$ is the $d$-dimensional vector of ones.
\end{Definition}
\begin{Definition}[Risk contributions and conditional risk measures]
  For $j \in \{1,\dots,d\}$, we call $A_j^{\varrho_j,\mathcal C}$ of
  \begin{enumerate}
  \item \emph{risk contribution type} if $\varrho_j=\E$;
  \item \emph{CoVaR type} if  $\varrho_j=\operatorname{VaR}_{\beta_j}$ for $\beta_j \in (0,1)$;
  \item \emph{CoRVaR type} if  $\varrho_j=\RVaR_{\beta_{j,1},\beta_{j,2}}$ for $0< \beta_{j,1} < \beta_{j,2} \leq 1$; and
  \item \emph{CoES type} if  $\varrho_j=\ES_{\beta_j}$ for $\beta_{j} \in (0,1)$.
  \end{enumerate}
\end{Definition}
The following examples show that $A_j^{\varrho_j,\mathcal C}$ coincides with
popular allocations for some specific choices of marginal risk measure and
crisis event.

\begin{Example}[Special cases of $A^{\sqc{\varrho}{d},\mathcal C}$]\mbox{}
  \begin{enumerate}
  \item Risk contributions. If the crisis event is chosen to be
    $\mathcal C_{\alpha}^{\operatorname{VaR}}$, $\mathcal C_{\alpha_1,\alpha_2}^{\RVaR}$
    or $\mathcal C_{\alpha}^{\ES}$, the allocations of the risk contribution
    type $\varrho_j=\E$ reduce to the \emph{VaR, RVaR or ES contributions}
    defined by
      \begin{align*}
      \VaR_\alpha(\bX,S)&= \E[\bX \ | \  S = \VaR_{\alpha}(S)],\\
      \RVaR_{\alpha_1,\alpha_2}(\bX,S)&= \E[\bX \ | \  \VaR_{\alpha_1}(S)\leq S \leq \VaR_{\alpha_2}(S)],\\
      \ES_\alpha(\bX,S)&= \E[\bX \ | \  S \geq  \VaR_{\alpha}(S)],
    \end{align*}
    respectively.  These results are derived by allocating the total capital
    $\VaR_{\alpha}(S)$, $\RVaR_{\alpha_1,\alpha_2}(S)$ and $\ES_{\alpha}(S)$
    according to the Euler principle; see \citet{tasche1999risk}.  The ES
    contribution is also called the MES and
    used as a systemic risk measure; see \citet{acharya2017measuring}.
  \item Conditional risk measures. CoVaR and CoES are systemic risk measures defined by
     \begin{align*}
      \CoVaR_{\alpha,\beta}^{=}(X_j,S)&=\VaR_{\beta}(X_j|S = \VaR_{\alpha}(S)),\quad
                                        \CoVaR_{\alpha,\beta}(X_j,S)=\VaR_{\beta}(X_j|S \geq  \VaR_{\alpha}(S)),\\
      \CoES_{\alpha,\beta}^{=}(X_j,S)&=\ES_{\beta}(X_j|S = \VaR_{\alpha}(S)),\quad
                                       \CoES_{\alpha,\beta}(X_j,S)=\ES_{\beta}(X_j|S \geq \VaR_{\alpha}(S)),
    \end{align*}
    for $\alpha, \beta \in (0,1)$; see \citet{mainik2014dependence} and \citet{bernardi2017covar}.  Our CoVaR and CoES type allocations with
    crisis events $\mathcal C =\mathcal C^{\operatorname{VaR}_{\alpha}}$ or $\mathcal C^{\ES_{\alpha}}$ coincide with
    those defined in the last displayed equations.
  \end{enumerate}
\end{Example}

\begin{Remark}[Weighted allocations]
  For a measurable function $w:\IR^d \rightarrow \IR_{+}:=[0,\infty)$,
  \citet{furman2008weighted} proposed the \emph{weighted allocation}
  $\varrho_w(\bX)$ with the \emph{weight function} $w$ being defined by
  $\varrho_w(\bX)=\E[\bX w(\bX)]/\E[w(\bX)]$.  By taking an indicator function
  as weight function $w(\bx)= \bone_{[\bx\in \mathcal C]}$ and provided that
  $\Prob(\bX \in \mathcal C)>0$, weighted allocation coincides with the
  risk contribution type systemic allocation $A^{\E,\dots,\E, \mathcal C}$.
\end{Remark}

\subsection{Monte Carlo Estimation of Systemic Risk Allocations}\label{subsec:MC:estimation:allocation}
Even if the joint distribution $F_{\bX}$ of the loss random vector $\bX$ is
known, the conditional distribution of $\X$ given $\bX \in \mathcal C$, denoted
by $F_{\bX|\mathcal C}$, is typically too complicated to analytically calculate
the systemic risk allocations $A^{\varrho_1,\dots,\varrho_d,\mathcal C}$.  An alternative
approach is to numerically estimate them by the MC
method as is done in \citet{yamai2002comparative} and \citet{fan2012decomposition}.  To this end, assume that one can generate
i.i.d. samples from $F_{\bX}$.
If $\Prob(\bX \in \mathcal C) > 0$, the MC estimator of  $A_j^{\varrho_j,\mathcal C}$, $j=1,\dots,d$ is constructed as follows:
\begin{enumerate}
\item \emph{Sample from $\bX$}: For a sample size $N\in \IN$, generate $\bX^{(1)},\dots,\bX^{(N)}\isim F_{\bX}$.
\item \emph{Estimate the crisis event}: If the crisis event $\mathcal C$ contains unknown quantities, replace them with their estimates based on $\bX^{(1)},\dots,\bX^{(N)}$.
  Denote by $\hat{\mathcal C}$ the estimated crisis event.
\item \emph{Sample from the conditional distribution of $\bX$ given $\hat{\mathcal C}$}: Among $\bX^{(1)},\dots,\bX^{(N)}$, determine $\tilde \bX^{(n)}$ such that $\tilde \bX^{(n)} \in \hat{\mathcal C}$ for all $n=1,\dots,N$.
\item \emph{Construct the MC estimator}: The MC estimate of $A_j^{\varrho_j,\mathcal C}$ is $\varrho_j(\hat F_{\tilde \bX})$ where $\hat F_{\tilde \bX}$ is the empirical cdf (ecdf) of the $\tilde \bX^{(n)}$'s.
\end{enumerate}
For an example of (2), if the crisis event is $\mathcal C^{\RVaR_{\alpha_1,\alpha_2}}= \{\bx \in \IR^d\ | \ \VaR_{\alpha_1}(S)\leq \bone_d\T \bx \leq \VaR_{\alpha_2}(S) \}$, then $\VaR_{\alpha_1}(S)$ and $\VaR_{\alpha_2}(S)$ are unknown parameters, and thus they are replaced by $\VaR_{\alpha_1}(\hat{F}_S)$ and $\VaR_{\alpha_2}(\hat{F}_S)$, where $\hat{F}_S$ is the ecdf of the total loss $S^{(n)}:=X_{1}^{(n)}+\cdots+X_{d}^{(n)}$ for $n=1,\dots,N$.
By the \emph{law of large numbers (LLN)} and the \emph{central limit theorem (CLT)}, the MC estimator of $A^{\sqc{\varrho}{d},\mathcal C}$ is consistent, and approximate confidence interval of the true allocation can be constructed based on asymptotic normality; see \citet{glasserman2005measuring}.

MC cannot handle VaR crisis events if $S$ admits a pdf since
$\Prob(\bX \in \mathcal C^{\operatorname{VaR}_{\alpha}})=\Prob(S = \VaR_{\alpha}(S))=0$
and thus no subsample is picked in (3) above.  A possible remedy (although the
resulting estimator suffers from an inevitable bias) is to replace
$\mathcal C_{\alpha}^{\operatorname{VaR}}$ with
$\mathcal C_{\alpha-\delta,\alpha+\delta}^{\RVaR}$ for sufficiently small
$\delta >0$ so that
$\Prob(S \in \mathcal C_{\alpha-\delta,\alpha+\delta}^{\RVaR})=2\delta > 0$.

The main advantage of MC for estimating systemic risk allocations
$A^{\sqc{\varrho}{d},\mathcal C}$ is that only a random number generator for
$F_{\bX}$ is required for implementing the method.  Furthermore, MC is
applicable for any choice of the crisis event $\mathcal C$ as long as
$\Prob(\bX \in \mathcal C)>0$.  Moreover, the main computational load is
simulating $F_{\bX}$ in (1) above, which is typically not demanding. The
disadvantage of the MC method is its inefficiency concerning the rare-event
characteristics of $\sqc{\varrho}{d}$ and $\mathcal C$.  To see this, consider
the case where $\mathcal C= \mathcal C_{\alpha_1,\alpha_2}^{\RVaR}$ and
$\varrho_{j}=\RVaR_{\beta_1,\beta_2}$ for $\alpha_1=\beta_1= 0.95$ and
$\alpha_2=\beta_2 =0.975$.  If the MC sample size is $N=10^5$, there are $N\times (\alpha_2-\alpha_1) = 2500$ subsamples resulting from (3).
To estimate RVaR$_{\beta_1,\beta_2}$ in (4) based on this subsample, only
$2500\times (\beta_2-\beta_1) = 62.5$ samples contribute to
computing the estimate, which is in general not enough for statistical
inference. This effect of sample size reduction is relaxed if ES and/or ES
crisis events are considered, but is more problematic for the VaR crisis
event since there is a trade-off concerning reducing bias and MC error when
choosing $\delta$; see \citet{koike2019estimation}.

\section{MCMC Estimation of Systemic Risk Allocations}\label{sec:mcmc:estimation:systemic:risk:allocation}
To overcome the drawback of the MC method for estimating systemic risk allocations, we introduce MCMC methods which simulate a given distribution by constructing a Markov chain whose stationary distribution is $F_{\bX|\mathcal C}$.
In this section, we first briefly review MCMC methods including the MH algorithm as a major subclass of MCMC methods, and then study how to construct an efficient MCMC estimator for the different choices of crisis events.

\subsection{A Brief Review of MCMC}\label{subsec:brief:review:MCMC}
Let $E\subseteq \IR^{d}$ be a set and $\mathcal E$ be a $\sigma$-algebra on $E$.
A \emph{Markov chain} is a sequence of $E$-valued random variables
$(\bX^{(n)})_{n \in \IN_0}$ satisfying the Markov property $\Prob(\bX^{(n+1)}\in A\ |\ \bX^{(k)}=\bx^{(k)},k\leq n)=\Prob(\bX^{(n+1)}\in A\ |\ \bX^{(n)}=\bx^{(n)})$
for all $n\geq 1$, $A \in \mathcal E$, and $\bx^{(1)},\dots,\bx^{(n)}\in E$.  A
Markov chain is characterized by its \emph{stochastic kernel}
$K: E \times \mathcal E \rightarrow [0,1]$ given by
$\bx \times A \mapsto K(\bx,A):=\Prob(\bX^{(n+1)}\in A\ |\ \bX^{(n)}=\bx)$.  A
probability distribution $\pi$ satisfying
$\pi(A)=\int_{E} \pi(\text{d}\bx) K(\bx,A)$ for any $\bx\in E$ and
$A \in \mathcal E$ is called \emph{stationary distribution}.  Assuming
$K(\bx, \cdot)$ has a density $k(\bx,\cdot)$, the \emph{detailed balance
  condition} (also known as the \emph{reversibility}) with respect to $\pi$ is
given by
\begin{align}
  \pi(\bx)k(\bx,\by) = \pi(\by)k(\by,\bx), \quad \bx,\by \in E,\label{eq:detailed:balance:condition}
\end{align}
and is known as a sufficient condition for the corresponding kernel $K$ to have
the stationary distribution $\pi$; see \citet{chib1995understanding}.
MCMC methods simulate a distribution as a
sample path of a Markov chain whose stationary distribution $\pi$ is the desired
one.  For a given distribution $\pi$, also known as \emph{target distribution},
and a functional $\varrho$, the quantity of interest $\varrho(\pi)$ is estimated
by the MCMC estimator $\varrho(\hat{\pi})$ where $\hat{\pi}$ is the empirical
distribution constructed from a sample path
$\bX^{(1)},\dots,\bX^{(N)}$ of the Markov chain whose stationary distribution is $\pi$.  Under
regularity conditions, the MCMC estimator is consistent and asymptotically
normal; see \citet{nummelin2002mc}, \citet{nummelin2004general} and
\citet{meyn2012markov}.  Its asymptotic variance can be estimated from
$(\bX^{(1)},\dots,\bX^{(N)})$ by, for instance, the \emph{batch
  means estimator}; see \citet{jones2006fixed}, \citet{geyer2011introduction} 
and \citet{vats2015multivariate} for more details.  Consequently, one can
construct approximate confidence intervals for the true quantity $\varrho(\pi)$
based on a sample path of the Markov chain.


Since the target distribution $\pi$ is determined by the problem at hand, the
problem is to find the stochastic kernel $K$ having $\pi$ as stationary
distribution such that the corresponding Markov chain can be easily simulated.
One of the most prevalent stochastic kernels is the \emph{Metropolis-Hastings
  (MH) kernel} defined by
$K(\bx,{\rm d}\by) = k(\bx,\by){\rm d}\by + r(\bx)\delta_{\bx}(\rd \by)$, where
$\delta_{\bx}$ is the Dirac delta function;
$k(\bx,\by)=q(\bx,\by)\alpha(\bx,\by)$; $q: E\times E \rightarrow \IR_{+}$ is a
function called a \emph{proposal density} such that $\bx \mapsto q(\bx,\by)$ is
measurable for any $\by \in E$ and $\by \mapsto q(\bx,\by)$ is a probability
density for any $\bx \in E$;
\begin{align*}
  \alpha(\bx,\by) = \begin{cases}
    \text{min}\left\{ \frac{\pi(\by)q(\by,\bx)}{\pi(\bx)q(\bx,\by)}, 1 \right\},&\text{ if } \pi(\bx)q(\bx,\by)>0,\\
    0,&\text{ otherwise};
  \end{cases}
\end{align*}
and $r(\bx) = 1- \int_{E} k(\bx, \by ){\rm d}\by$.
It can be shown that the MH kernel has stationary distribution $\pi$; see \citet{tierney1994markov}.
Simulation of the Markov chain with this MH kernel is conducted by the \emph{MH algorithm} given in Algorithm~\ref{algorithm:MH}.
\begin{algorithm}[t]
  \caption{Metropolis-Hastings (MH) algorithm}
  \label{algorithm:MH}
  \begin{algorithmic}[0]
    \REQUIRE Random number generator of the proposal density $q(\bx,\cdot)$ for all $\bx \in E$, $\bx^{(0)}\in \text{supp}(\pi)$ and the ratio $\pi(\by)/\pi(\bx)$ for $\bx,\by \in E$ where $\pi$ is the density of the stationary distribution.\\
    \INPUT Sample size $N\in \IN$, proposal density $q$, and initial value $\bm X^{(0)}=\bx^{(0)}$.\\
    \OUTPUT Sample path $\bX^{(1)},\dots,\bX^{(N)}$ of the Markov chain.\\ \vspace{1mm}
    \algrule
    \FOR{$n:=0,\dots,N-1$}\\
    \STATE 1) Generate $\tilde \bX^{(n)}\sim q(\bX^{(n)},\cdot)$\\
    \STATE 2) Calculate the \emph{acceptance probability}
    \begin{align}
      \alpha_{n}:=\alpha(\bX^{(n)},\tilde \bX^{(n)})=\min \left\{
        \frac{\pi(\tilde \bX^{(n)})q(\tilde \bX^{(n)},\bX^{(n)})}{\pi(\bX^{(n)})q(\bX^{(n)},\tilde \bX^{(n)})},
        1\right\}.\label{eq:acceptance:probability}
    \end{align}
    \STATE 3) Generate $U\sim \mathcal U(0,1)$ and set $\bX^{(n+1)}:=1_{[U\leq \alpha_{n}]} \tilde \bX^{(n)}+ 1_{[U>\alpha_{n}]} \bX^{(n)}$.\\
    \ENDFOR
  \end{algorithmic}
\end{algorithm}

An advantage of the MCMC method is that a wide variety of distributions can be
simulated as a sample path of a Markov chain even if generating i.i.d. samples
is not feasible directly.  The price to pay is an additional computational cost
to calculate the acceptance probability~\eqref{eq:acceptance:probability}, and a
possibly higher standard deviation of the estimator $\varrho(\hat{\pi})$
compared to the standard deviation of estimators constructed from
i.i.d. samples.  This attributes to the serial dependence among MCMC
samples which can be seen as follows. Suppose first that the candidate
$\tilde \bX^{(n)}$ is rejected (so $\{U>\alpha_{n}\}$ occurs).  Then
$ \bX^{(n+1)}=\bX^{(n)}$ and thus the samples are perfectly dependent.  The candidate $\tilde \bX^{(n)}$ is more likely to be accepted if
the acceptance probability $\alpha_n$ is close to 1.  In this case,
$\pi(\bX^{(n)})$ and $\pi(\tilde \bX^{(n)})$ are expected to be close to each
other (otherwise $\pi(\tilde \bX^{(n)})/\pi(\tilde \bX^{(n)})$ and thus
$\alpha_n$ can be small).  Under the continuity of $\pi$, $\tilde \bX^{(n)}$ and
$\bX^{(n)}$ are expected to be close and thus dependent with each other.
An efficient MCMC method is such that the candidate $\tilde \bX^{(n)}$ is
sufficiently far from $\bX^{(n)}$ with the probability $\pi(\tilde \bX^{(n)})$
as close to $\pi(\bX^{(n)})$ as possible.  Efficiency of MCMC can indirectly be
inspected through the \emph{acceptance rate (ACR)} and the \emph{autocorrelation
  plot (ACP)}; ACR is the percentage of times a candidate $\tilde \bX$ is
accepted among the $N$ iterations, and ACP is the plot of the autocorrelation
function of the generated sample path.  An efficient MCMC method shows high ACR
and steady decline in ACP; see \citet{chib1995understanding} and
\citet{rosenthal2011optimal} for details.  Ideally the proposal density $q$ is
constructed only based on $\pi$ but typically $q$ is chosen among a parametric
family of distributions.  For such cases, simplicity of the choice of tuning
parameters of $q$ is also important.

\subsection{MCMC Formulation for Estimating Systemic Risk Allocations}\label{subsec:mcmc:formulation:estimating:systemic:allocations}
Numerous choices of proposal densities $q$ are possible to construct an MH kernel.
In this subsection, we consider how to construct an efficient MCMC method for
estimating systemic risk allocations $A^{\sqc{\varrho}{d},\mathcal C}$ depending
on the choice of the crisis event $\mathcal C$.  Our goal is to simulate the
conditional distribution $\bX|\mathcal C$ directly by constructing a Markov
chain whose stationary distribution is
\begin{align}
  \pi(\bx)=f_{\bX| \bX \in \mathcal C}(\bx)= \frac{f_{\bX}(\bx)}{\Prob(\bX \in \mathcal C)}\bone_{[\bx \in \mathcal C]},\quad \bx \in E\subseteq \IR^d,\label{eq:target:distribution:conditional}
\end{align}
provided $\Prob(\bX \in \mathcal C)> 0$.  Samples from this distribution can
directly be used to estimate systemic risk allocations with crisis event
$\mathcal C$ and arbitrary marginal risk measures $\sqc{\varrho}{d}$.  Other
potential applications are outlined in
Remark~\ref{remark:potential:applications:conditional:samples}.
\begin{Remark}[Gini shortfall allocation]\label{remark:potential:applications:conditional:samples}
  Samples from the conditional distribution $F_{\bX|\mathcal C_{\alpha}^{\ES}}$ can be used to
  estimate, for example, the \emph{tail-Gini coefficient}
  $\text{TGini}_{\alpha}(X_j,S)=\frac{4}{1-\alpha}\Cov(X_j,F_S(S)\ |\ S \geq
  \operatorname{VaR}_{\alpha}(S))$ for $\alpha \in (0,1)$, and the \emph{Gini shortfall
    allocation} (\citet{furman2017gini})
  $\text{GS}_{\alpha}(X_j,S)=\E[X_j\ | \ S \geq \operatorname{VaR}_{\alpha}(S)]+\lambda
  \cdot \text{TGini}_{\alpha}(X_j,S)$, $\lambda \in \IR_{+}$ more efficiently
  than by applying the MC method.
  Another application is to estimate risk
  allocations derived by optimization given a constant economic capital; see
  \citet{laeven2004optimization} and \citet{dhaene2012optimal}.
\end{Remark}

We now construct a MH algorithm with target
distribution~\eqref{eq:target:distribution:conditional}.  To this end, we assume that
\begin{enumerate}
\item \label{eq:Assumption:ratio:computable}
  \text{the ratio $f_{\bX}(\by)/f_{\bX}(\bx)$ can be evaluated for any $\bx,\by \in \mathcal C$}, and that
\item \label{eq:Assumption:support}
  \text{the support of $f_{\bX}$ is $\IR^d$ or $\IR_{+}^d$}.
\end{enumerate}
Regarding Assumption~\ref{eq:Assumption:ratio:computable}, the normalization
constant of $f_{\bX}$ and the probability $\Prob(\bX \in \mathcal C)$ are not
necessary to be known since they cancel out in the numerator and the denominator
of $\pi(\by)/\pi(\bx)$.  In Assumption~\ref{eq:Assumption:support},
the loss random vector $\bX$ refers to the \emph{profit$\&$loss (P$\&$L)} if
$\supp{f_{\bX}}=\IR^d$ and to \emph{pure losses} if
$\supp{f_{\bX}}=\IR_{+}^d$.  Note that the case
$\supp{f_{\bX}}=[c_1,\infty]\times \cdots \times [c_d,\infty]$,
$\sqc{c}{d} \in \IR$ is essentially included in the case of pure losses
as long as the marginal risk measures $\sqc{\varrho}{d}$ are law invariant and
translation invariant, and the crisis event is the set of linear constraints of
Form~\eqref{eq:crisis:event:intersections}.  To see this, define
$\tilde X_j=X_j-c_j$, $j=1,\dots,d$, $\tilde \bX=(\sqc{\tilde X}{d})$
and $\bc = (\sqc{c}{d})$.  Then $\supp{f_{\tilde \bX}}=\IR_{+}^d$ and
$\bX | (\bX \in \mathcal C)\deq \tilde \bX|(\tilde \bX \in \tilde {\mathcal C})
+ \bc$ where $\tilde {\mathcal C}$ is the set of linear constraints with
parameters $\tilde \bh_m = \bh_m$ and $\tilde v_m = v_m - \bh_m\T\bc$.  By law
invariance and translation invariance of $\sqc{\varrho}{d}$,
\begin{align*}
  \varrho_j(X_j | \bX \in \mathcal C) = c_j + \varrho_j(\tilde X_j | \tilde \bX \in \tilde{\mathcal C}),\quad j=1,\dots,d.
\end{align*}
Therefore, the problem of estimating $A^{\sqc{\varrho}{d},\mathcal C}(\bX)$
reduces to that of estimating
$A^{\sqc{\varrho}{d},\tilde{\mathcal C}}(\tilde \bX)$ for the shifted loss
random vector $\tilde \bX$ (such that $\supp{f_{\tilde \bX}}=\IR_{+}^d$) and the
modified crisis event of the same form.

For the P$\&$L case, the RVaR and ES crisis events are the
set of linear constraints of Form~\eqref{eq:crisis:event:intersections} with the
number of constraints $M=2$ and $1$, respectively.  In the case of pure losses,
additional $d$ constraints $\be_{j,d}\T\bx \geq 0$, $j=1,\dots,d$ are imposed
where $\be_{j,d}$ is the $j$th $d$-dimensional unit vector.  Therefore, the RVaR
and ES crisis events are of Form~\eqref{eq:crisis:event:intersections} with
$M=d+2$ and $d+1$, respectively.  For the VaR crisis event,
$\Prob(\bX \in \mathcal C)= 0$ and thus
\eqref{eq:target:distribution:conditional} cannot be properly defined.  In this
case, the allocation $A^{\sqc{\varrho}{d},\mathcal C^{\operatorname{VaR}}}$ depends on
the conditional joint distribution $\bX|\mathcal C_{\alpha}^{\operatorname{VaR}}$ but is
completely determined by its first $d' := d-1$ variables
$(X_1,\dots,X_{d'})|\mathcal C_{\alpha}^{\operatorname{VaR}}$ since
$\smash{X_d | \mathcal C_{\alpha}^{\operatorname{VaR}} \deq (\VaR_{\alpha}(S) -
\sum_{j=1}^{d'}X_j) | \mathcal C_{\alpha}^{\operatorname{VaR}} \deq \VaR_{\alpha}(S) -
\sum_{j=1}^{d'} X_j | \mathcal C_{\alpha}^{\operatorname{VaR}}}$.  Estimating systemic
risk allocations under the VaR crisis event can thus be achieved by simulating
the target distribution
\begin{align}
  \pi^{\operatorname{VaR}_\alpha}(\bx')&=f_{\bX'| S =\VaR_{\alpha}(S)}(\bx)= \frac{f_{(\bX',S)}(\bx',\VaR_{\alpha}(S))}{f_{S}(\VaR_{\alpha}(S))}\notag\\
  &=\frac{f_{\bX}(\bx',\VaR_{\alpha}(S)-\bone_{d'}\T\bx')}{f_{S}(\VaR_{\alpha}(S))}\bone_{[\VaR_{\alpha}(S)-\bone_{d'}\T\bx' \in \text{supp}(f_d)]}, \quad \bx' \in \IR^{d'},\label{eq:VaR:crisis:target:distribution:conditional}
\end{align}
where $\bX'=(\sqc{X}{d'})$ and the last equation is derived from the linear
transformation $(\bX',S)\mapsto \bX$ with unit Jacobian.  Note that other
transformations are also possible; see \citet{betancourt2012cruising}. Under
Assumption~\ref{eq:Assumption:ratio:computable}, the ratio
$\pi^{\operatorname{VaR}_{\alpha}}(\by)/\pi^{\operatorname{VaR}_{\alpha}}(\bx)$ can be evaluated
and $f_{S}(\VaR_{\alpha}(S))$ is not required to be known.  In the case of pure
losses, the target distribution $\pi^{\operatorname{VaR}_\alpha}$ is subject to $d$
linear constraints $\be_{j,d'}\T\bx' \geq 0$, $j=1,\dots,d'$ and
$\bone_{d'}\T\bx' \geq \VaR_{\alpha}(S)$ where the first $d'$ constraints come
from the non-negativity of the losses and the last one is from the indicator
in~\eqref{eq:VaR:crisis:target:distribution:conditional}. Therefore, the crisis
event $\mathcal C^{\operatorname{VaR}}$ for $(\sqc{X}{d'})$ is of 
Form~\eqref{eq:crisis:event:intersections}.  In the case of P$\&$L,
$\text{supp}(f_d) = \IR$ and
$\VaR_{\alpha}(S)-\bone_{d'}\T\bx' \in \text{supp}(f_d)$ holds for any
$\bx' \in \IR^{d'}$.  Therefore, the target
distribution~\eqref{eq:VaR:crisis:target:distribution:conditional} is free from
any constraints and the problem reduces to constructing an MCMC method with
target distribution
$\pi(\bx')\propto f_{\bX}(\bx', \VaR_{\alpha}(S)-\bone_{d'}\T\bx')$,
$\bx' \in \IR^{d'}$.  In this paper the P$\&$L case with VaR crisis event is not
investigated further since our focus is the simulation of constrained
target distributions; see \citet{koike2019estimation} for an MCMC estimation in
the P$\&$L case.

MCMC methods to simulate constrained target distributions require careful design
of the proposal density $q$.  A simple MCMC method is \emph{Metropolis-Hastings
  with rejection} in which the support of the proposal density $q$ may not
coincide with that of the target distribution, which is the crisis event $\mathcal C$, and a candidate is immediately rejected when it
violates the constraints.  This construction of MCMC is often inefficient due to
a low acceptance probability especially around the bondary of $\mathcal C$.  An
efficient MCMC method in this case can be expected only when the probability
mass of $\pi$ is concentrated near the center of $\mathcal C$.  In the following
sections, we introduce two alternative MCMC methods for the constrained target
distributions $F_{\bX|\mathcal C}$ of interest, the HMC
method and the GS.  Each of them is applicable and can be efficient
for different choices of the crisis event and underlying loss distribution
functions $F_{\bX}$.

\subsection{Estimation with Hamiltonian Monte Carlo}\label{subsec:HMC}
We find that if the HMC method is applicable, it is typically the most
preferable method to simulate constrained target distributions because of its
efficiency and ease of handling constraints.  In
Section~\ref{subsubsec:HMC:reflection}, we briefly present the HMC method with
reflection for constructing a Markov chain supported on the constrained space.
In Section~\ref{subsubsec:choice:parameters:HMC} we propose a heuristic
  for determining the parameters of the HMC method based on the MC presamples.

\subsubsection{Hamiltonian Monte Carlo with Reflection}\label{subsubsec:HMC:reflection}
For the possibly unnormalized target density $\pi$, consider the \emph{potential
  energy} $U(\bx)$, \emph{kinetic energy} $K(\bp)$ and the \emph{Hamiltonian}
$H(\bx,\bp)$ defined by 
  \begin{align*}
    U(\bx)=-\log\pi(\bx),\quad K(\bp)= -\log f_K(\bp)\quad\text{and}\quad H(\bx,\bp) = U(\bx)+K(\bp),
  \end{align*}
with \emph{position variable} $\bx \in E$, \emph{momentum variable}
$\bp\in \IR^d$ and \emph{kinetic energy density} $f_K(\bp)$ such that
$f_K(-\bp)=f_K(\bp)$.  In this paper, the kinetic energy
distribution $F_K$ is set to be the multivariate standard normal with
$K(\bp)=\frac{1}{2}\bp \T \bp$ and $\nabla K(\bp)= \bp$; other choices of $F_K$
are discussed in Appendix~\ref{appendix:subsec:riemannian:manifold:HMC}.  In the HMC
method, a Markov chain augmented on the state space $E\times \IR^d$ with the
stationary distribution $\pi(\bx)f_K(\bp)$ is constructed and the desired
samples from $\pi$ are obtained as the first $|E|$-dimensional margins.  A
process $(\bx(t),\bp(t))$, $ t \in \IR$ on $E\times \IR^d$ is said to follow the
\emph{Hamiltonian dynamics} if it follows the ordinary differential equation
(ODE)
\begin{align}
  \frac{\rd}{\rd t}\bx(t) = \nabla K(\bp),\quad
  \frac{\rd}{\rd t}\bp(t) = -\nabla U(\bx).\label{eq:hamiltonian:dynamics}
\end{align}
Through the Hamiltonian dynamics, the Hamiltonian $H$ and the volume are
conserved, that is, 
$\rd H(\bx(t),\bp(t))/\rd t=0$ and the map
$(\bx(0),\bp(0))\mapsto (\bx(t),\bp(t))$ has a unit Jacobian for any
$ t \in \IR$; see \citet{neal2011mcmc}.  Therefore, the value of the joint target density $\pi \cdot f_K$ remains unchanged by the
Hamiltonian dynamics, that is,
  \begin{align*}
    \pi(\bx(0))f_K(\bp(0)) = \exp(- H(\bx(0),\bp(0))) = \exp(- H(\bx(t),\bp(t))) = \pi(\bx(t))f_K(\bp(t)),\quad t \geq 0.
  \end{align*}
In practice, the dynamics~\eqref{eq:hamiltonian:dynamics} are discretized for simulation by, for example, the so-called \emph{leapfrog method} summarized in Algorithm~\ref{algorithm:LF}; see \citet{leimkuhler2004simulating} for other discretization methods.
\begin{algorithm}[t]
  \caption{Leapfrog method for Hamiltonian dynamics}
  \label{algorithm:LF}
  \begin{algorithmic}[0]
    \INPUT  Current states $(\bx(0),\bp(0))$, stepsize $\epsilon>0$, gradients $\nabla U$ and $\nabla K$.\\
    \OUTPUT Updated position $(\bx(\epsilon),\bp(\epsilon))$. \\ \vspace{1mm}
    \algrule
    \STATE 1) $\bp\left(\frac{\epsilon}{2}\right) = \bp(0) - \frac{\epsilon}{2} \nabla U(\bx(0))$.\\
    \STATE 2) $\bx(\epsilon) = \bx(0) + \epsilon \nabla K(\bp\left(\frac{\epsilon}{2}\right))$.\\
    \STATE 3) $\bp(\epsilon) = \bp(\epsilon/2) + \frac{\epsilon}{2} \nabla U(\bx(\epsilon))$.\\
  \end{algorithmic}
\end{algorithm}
Note that the evaluation of $\nabla U$ does not require the normalization
constant of $\pi$ to be known since $\nabla U = - (\nabla \pi)/\pi$.  By
repeating the leapfrog method $T$ times with stepsize $\epsilon$, the
Hamiltonian dynamics are approximately simulated with length $T\epsilon$.  Due
to the discretization error the Hamiltonian is not exactly preserved while it is
expected to be almost preserved for $\epsilon$ small enough.
The discretization error $H(\bx(T\epsilon),\bp(T\epsilon))-H(\bx(0),\bp(0))$ is called the \emph{Hamiltonian error}.

All the steps of the HMC method are described in Algorithm~\ref{algorithm:HMC}.
In Step 1), the momentum variable is first updated from $\bp(0)$ to $\bp$ where
$\bp$ follows the kinetic energy distribution $F_K$ so that the value of the
Hamiltonian $H= -\log (\pi \cdot f_K)$ changes.  In Step 3), the current
state $(\bx(0),\bp)$ is moved along the level curve of $H(\bx(0),\bp)$ by
simulating the Hamiltonian dynamics.
\begin{algorithm}[t]
  \caption{Hamiltonian Monte Carlo to simulate $\pi$}
  \label{algorithm:HMC}
  \begin{algorithmic}[0]
    \REQUIRE Random number generator of $F_K$, $\bx^{(0)}\in \text{supp}(\pi)$, $\pi(\by)/\pi(\bx)$, $\bx,\by \in E$ and $f_K(\bp')/f_K(\bp)$, $\bp,\bp' \in \IR^d$. \\
    \INPUT Sample size $N\in \IN$, kinetic energy density $f_K$, target density $\pi$, gradients of the potential and kinetic energies $\nabla U$ and   $\nabla K$, stepsize $\epsilon>0$, integration time $T\in \IN$ and initial position $\bm X^{(0)}=\bx^{(0)}$.\\
    \OUTPUT Sample path $\bX^{(1)},\dots,\bX^{(N)}$ of the Markov chain.\\  \vspace{1mm}
    \algrule
    \FOR{$n:=0,\dots,N-1$}\\
    \STATE 1) Generate $\bp^{(n)} \sim F_K$.\\
    \STATE 2) Set $(\tilde \bX^{(n)}, \tilde \bp^{(n)}) = (\bX^{(n)}, \bp^{(n)})$.\\
    \STATE 3) \textbf{for} $t:=1,\dots,T$,
    \begin{align*}
      (\tilde \bX^{(n+t/T)}, \tilde \bp^{(n+t/T))} =\ \text{\textbf{Leapfrog}}(\tilde \bX^{(n+(t-1)/T)}, \tilde \bp^{(n+(t-1)/T)}, \epsilon, \nabla U, \nabla K).
    \end{align*}\\
    \STATE \hspace{3mm} \textbf{end for} \\
    \STATE 4) $\tilde \bp^{(n+1)} = -\bp^{(n+1)}$.\\
    \STATE 5) Calculate
    $\alpha_{n}=\min \left\{
      \frac{\pi(\tilde \bX^{(n+1)})f_K(\tilde \bp^{(n+1)})}{\pi(\bX^{(n)})f_K(\bp^{(n)})},
      1\right\}$.\\ \vspace{3mm}
    \STATE 6) Set $\bX^{(n+1)}:=1_{[U\leq \alpha_{n}]} \tilde \bX^{(n+1)}+ 1_{[U>\alpha_{n}]} \bX^{(n)}$ for $U \sim \Unif(0,1)$.\\
    \ENDFOR
  \end{algorithmic}
\end{algorithm}
By flipping the momentum in Step 4), the HMC method is shown to be reversible w.r.t.\ $\pi$ (c.f.~\eqref{eq:detailed:balance:condition}) and thus to have the stationary distribution $\pi$; see \citet{neal2011mcmc} for details.
Furthermore, by the conservation property of the Hamiltonian dynamics, the acceptance probability in Step 5) is expected to be close to $1$.
Moreover, by taking $T$ sufficiently large, the candidate $\tilde \bX^{(n+1)}$ is expected to be sufficiently decorrelated from the current position $\bX^{(n)}$.
Consequently, the resulting Markov chain is expected to be efficient.

The remaining challenge for applying the HMC method to our problem of estimating
systemic risk allocations is how to handle the constraint $\mathcal C$.  As we
have seen in Section~\ref{subsec:class:systemic:allocations}
and~\ref{subsec:mcmc:formulation:estimating:systemic:allocations}, $\mathcal C$
is assumed to be an intersection of linear constraints with parameters
$(\bh_m,v_m)$, $m=1,\dots,M$ describing hyperplanes.  Following the ordinary
leapfrog method, a candidate is immediately rejected when the trajectory of the
Hamiltonian dynamics penetrates one of these hyperplanes. To avoid it, we modify the leapfrog method according to the reflection
  technique introduced in \citet{afshar2015reflection} and
  \citet{chevallier2018hamiltonian}.  As a result, the trajectory is reflected
  when it hits a hyperplane and the Markov chain moves within the constrained
  space with probability one.  Details of the HMC method with reflection for our
  application are described in Appendix~\ref{appendix:reflection:technique:HMC}.

\subsubsection{Choice of Parameters for HMC}\label{subsubsec:choice:parameters:HMC}
HMC requires as input two parameters, the \emph{stepsize} $\epsilon$ and the \emph{integration
  time} $T$.  As we now explain, neither of them should be chosen too large nor
too small.  Since the stepsize $\epsilon$ controls the accuracy of the
simulation of the Hamiltonian dynamics, $\epsilon$ needs to be small enough to
approximately conserve the Hamiltonian; otherwise the acceptance probability can
be much smaller than $1$.  On the other hand, a too small $\epsilon$ requires
the integration time $T$ to be large for the trajectory to reach far, which is
computationally costly.  Next, the integration time $T$ needs to be large
enough to decorrelate the candidate state with the current state.  Meanwhile,
the trajectory of the Hamiltonian dynamics may make a U-turn and come back to the
starting point if the integration time $T$ is too long; see \citet{neal2011mcmc}
for an illustration of this phenomenon.

A notable characteristic of our problem of estimating systemic risk allocations
is that the MC sample from the target distribution $\pi$ is available but its
sample size may not be sufficient for statistical inference, and, in the case
of the VaR crisis event, the samples only approximately follow the target
distribution.  We utilize the information of this \emph{MC presample} to build a
heuristic for determining the parameters $(\epsilon,T)$; see
Algorithm~\ref{algorithm:adaptive:stepsize:integratio:time}.
\begin{algorithm}[t]
  \caption{Heuristic for determining the stepsize $\epsilon$ and integration time $T$}
  \label{algorithm:adaptive:stepsize:integratio:time}
  \begin{algorithmic}[0]
    \INPUT  MC presample $\bX_{1}^{(0)},\dots,\bX_{N_0}^{(0)}$, gradients $\nabla U$ and $\nabla K$, target acceptance probability $\underline \alpha$, initial constant $c_{\epsilon}>0$ and the maximum integration time $T_\text{max}$ ($c_{\epsilon}=1$ and $T_\text{max}=1000$ are set as default values).\\
    \OUTPUT Stepsize $\epsilon$ and integration time $T$.\\ \vspace{1mm}
    \algrule
    \STATE 1) Set $\alpha_\text{min}=0$ and $\epsilon = c_{\epsilon} d^{-1/4}$.\\
    \STATE 2) \textbf{while} $\alpha_\text{min}  <\underline \alpha$\\
    \STATE \hspace{8mm} 2-1) Set $\epsilon = \epsilon/2$.\\
    \STATE  \hspace{8mm} 2-2) \textbf{for} $n:=1,\dots,N_0$\\
    \STATE \hspace{8mm} \hspace{8mm} 2-2-1) Generate $\bp_n^{(0)} \sim F_K$.\\
    \STATE  \hspace{8mm} \hspace{8mm} 2-2-2) \textbf{for} $t:=1,\dots,T_\text{max}$\\
    \STATE \hspace{8mm} \hspace{8mm} \hspace{8mm} 2-2-2-1) Set $\bZ_n^{(t)}= \textbf{Leapfrog}(\bZ_n^{(t-1)},\epsilon,\nabla U,\nabla K)$ for $\bZ_n^{(t-1)} = (\bX_n^{(t-1)},\bp_n^{(t-1)})$.\\
    \STATE \hspace{8mm} \hspace{8mm} \hspace{8mm} 2-2-2-2) Calculate
    \begin{align*}
      \hspace{16mm}\alpha_{n,t}=\alpha(\bZ_n^{(t-1)},\bZ_n^{(t)})\quad\text{and}\quad
      \Delta_t = ||\bX_{n}^{(t)}-\bX_{n}^{(0)}|| - ||\bX_{n}^{(t-1)}-\bX_{n}^{(0)}||.
    \end{align*}\\
    \STATE  \hspace{8mm} \hspace{8mm} \hspace{8mm} 2-2-2-3) \textbf{if} $\Delta_{t}<0$ and $\Delta_{t-1} >0$, \textbf{break} and set $T_{n}^{\ast} =t-1$.\\
    \STATE  \hspace{8mm} \hspace{8mm}  \hspace{8mm} \textbf{end for}\\
    \STATE  \hspace{8mm} \hspace{6mm} \textbf{end for}\\
    \STATE \hspace{8mm} 2-3) Compute $\alpha_\text{min}=\min(\alpha_{n,t}\ | \ t=1,2,\dots,\ T_n^{\ast},\ n=1,\dots,N_0)$\\
    \STATE  \hspace{3mm} \textbf{end while}\\
    \STATE 3) Set $T = \lfloor \frac{1}{N_0}\sum_{n=1}^{N_0} T_{n}^{\ast} \rfloor$.\\
  \end{algorithmic}
\end{algorithm}
In this heuristic, the initial stepsize is set to be
$\epsilon = c_{\epsilon}d^{-1/4}$ for some constant $c_{\epsilon}>0$, say,
$c_{\epsilon}=1$.  This scale was derived in \citet{beskos2010acceptance} and
\citet{beskos2013optimal} under certain assumptions on the target distribution.
We determine $\epsilon$ through the relationship with the acceptance probability.  In Step 2-2-2-1) of
Algorithm~\ref{algorithm:adaptive:stepsize:integratio:time}, multiple
trajectories are simulated starting from each MC presample with the current
stepsize $\epsilon$.  In the next Step 2-2-2-2), we monitor the acceptance
probability and the distance between the starting and ending points while
extending the trajectories.  Based on the asymptotic optimal acceptance
probability 0.65 (c.f. \citet{gupta1990acceptance} and
\citet{betancourt2014optimizing}) as $d\rightarrow \infty$, we set the
\emph{target acceptance probability} as
\begin{align*}
  \underline \alpha = \frac{1 + (d-1)\times 0.65}{d}  \in (0.65,1].
\end{align*}
The stepsize is gradually decreased in Step 2-1) of
Algorithm~\ref{algorithm:adaptive:stepsize:integratio:time} until the minimum
acceptance probability calculated in Step 2-3) exceeds $\underline \alpha$.  To
prevent the trajectory from a U-turn, in Step 2-2-2-3) each trajectory is
immediately stopped when the distance begins to decrease.  The resulting
integration time is set to be the average of these turning points as seen in
Step 3).  Note that other termination conditions of extending trajectories are
possible; see \citet{hoffman2014no} and \citet{betancourt2016identifying}.

At the end of this section, we briefly revisit the choice of the kinetic energy
distribution $F_K$, which is taken to be multivariate standard normal throughout
this work.
As discussed in \citet{neal2011mcmc}, applying the HMC method with target
distribution $\pi$ and kinetic energy distribution $\N(\bzero,\Sigma^{-1})$ is
equivalent to applying HMC with the standardized target distribution
$\bx \rightarrow \pi(L\bx)$ and $F_K = \N(\bzero,\bm I)$ where $L$ is the
\emph{Cholesky factor} of $\Sigma$ such that $\Sigma= L L\T$.  By taking
$\Sigma$ to be the covariance matrix of $\pi$, the standardized target
distribution becomes uncorrelated with unit variances.  In our problem, the
sample covariance matrix $\hat \Sigma = \hat{L}\hat{L}\T$ calculated based on the
MC presample is used alternatively.  The new target distribution
$\tilde \pi(\by)=\pi(\hat L\by)|\hat L|$ where $|\hat L|$ denotes the Jacobian
of $\hat L$, is almost uncorrelated with unit variances, and thus the standard
normal kinetic energy fits well; see \citet{livingstone2019kinetic}.  If the
crisis event consists of the set of linear constraints $(\bh_m,v_m)$,
$m=1,\dots,M$, then the standardized target density is also subject to the set
of linear constraints $(\hat L \T \bh_m,v_m)$, $m=1,\dots,M$.  Since the ratio
$f_{\bX}(\hat L\by)/f_{\bX}(\hat L\bx)$ can still be evaluated under
Assumption~\ref{eq:Assumption:ratio:computable}, we conclude that the problem
remains unchanged after standardization.

Theoretical results of the HMC method with normal kinetic energy are available
only when $\mathcal C$ is bounded (\citet{cances2007theoretical} and
\citet{chevallier2018hamiltonian}), or when $\mathcal C$ is unbounded and the
tail of $\pi$ is roughly as light as that of the normal distribution
(\citet{livingstone2016geometric} and \citet{durmus2017convergence}).
Boundedness of $\mathcal C$ holds for VaR and RVaR crisis events with pure
losses; see \citet{koike2019estimation}.  As is discussed in this paper,
convergence results of MCMC estimators are accessible when the density of
the underlying joint loss distribution is bounded from above on $\mathcal C$, which is
typically the case when the underlying copula does not admit lower tail dependence.
For
other cases where $\mathcal C$ is unbounded or the density explodes on
$\mathcal C$, no convergence results are available.
Potential remedies for the HMC method to deal with heavy-tailed target distributions are discussed in Appendix~\ref{appendix:subsec:riemannian:manifold:HMC}.

\subsection{Estimation with Gibbs Sampler}\label{subsec:gibbs:samplers}
As discussed in Section~\ref{subsubsec:choice:parameters:HMC}, applying HMC
methods to heavy-tailed target distributions on unbounded crisis events is not
theoretically supported.  To deal with this case, we introduce the GS
in this section.

\subsubsection{True Gibbs Sampler for Estimating Systemic Risk Allocations}\label{subsubsec:GS:systemic:allocations}
The GS is a special case of the MH method in which the proposal density $q$
is completely determined by the target density $\pi$ via
\begin{align}
  q_{GS}(\bx,\by)= \sum_{\bbi = (\sqc{i}{d}) \in \mathcal I_d}p_{\bbi} \pi(y_{i_1}|\bx_{-i_1})\pi(y_{i_2}|y_{i_1},\bx_{-(i_1,i_2)})\cdots \pi(y_{i_d}|\by_{-i_d}),\label{eq:GS:proposal}
\end{align}
where $\bx_{-(j_1,\dots,j_l)}$ is the $(d-l)$-dimensional vector that excludes the components $j_1,\dots,j_l$ from $\bx$, $\pi(x_{j}|\bx_{-j})= \pi_{j|-j}(x_{j}|\bx_{-j})$ is the conditional density of the $j$th variable of $\pi$ given all the other components, $\mathcal I_d \subseteq \{1,\dots,d\}^d$ is the so-called \emph{index set} and $(p_{\bbi} \in [0,1],\ \bbi \in \mathcal I_d)$ is the \emph{index probability distribution} such that $\sum_{ \bbi \in \mathcal I_d}p_{\bbi}=1$.
For this choice of $q$, the acceptance probability is always equal to 1; see \citet{johnson2009geometric}.
The GS is called \emph{deterministic scan (DSGS)} if $\mathcal I_d= \{(1,\dots,d)\}$ and $p_{(1,\dots,d)}=1$.
When the index set is the set of permutations of $(1,\dots,d)$, the GS is called \emph{random permulation (RPGS)}.
Finally, the \emph{random scan GS (RSGS)} has the proposal~\eqref{eq:GS:proposal} with  $\mathcal I_d = \{1,\dots,d\}^d$ and $p_{(i_1,\dots,i_d)} = p_{i_1}\cdots p_{i_d}$ with probabilities $(\sqc{p}{d}) \in (0,1)^d$ such that $\sum_{j=1}^d p_j = 1$.
These three GSs can be shown to have $\pi$ as stationary distribution; see \citet{johnson2009geometric}.

Provided that the \emph{full conditional distributions} $\pi_{j|-j}$, $j=1,\dots,d$ can be simulated, the proposal distribution~\eqref{eq:GS:proposal} can be simulated by first selecting an index $\bbi \in \mathcal I_d$ with probability $p_{\bbi}$ and then replacing the $j$th component of the current state with a sample from $\pi_{j|-j}$ sequentially for $j=i_1,\dots,i_d$.
The main advantage of the GS is that the tails of $\pi$ are naturally incorporated via full conditional distributions, and thus the MCMC method is expected to be efficient even if $\pi$ is heavy-tailed.
On the other hand, the applicability of the GS is limited to target distributions such that $\pi_{j|-j}$ is available.
Moreover, fast simulators of $\pi_{j|-j}$, $j=1,\dots,d$, are required since the computational time linearly increases w.r.t.\ the dimension $d$.

In our problem of estimating systemic risk allocations, we find that the GS is applicable when the crisis event is of the form
\begin{align}
  \mathcal C = \{\bx \in \IR^d\text{ or }\IR_{+}^d\ | \ v_1 \leq \bh\T \bx \leq v_2 \},
  \quad v_1,v_2 \in \IR\cup \{\pm \infty\},
  \quad \bh = (\sqc{h}{d}) \in  \IR^d \backslash \{\bzero_d\}.\label{eq:crisis:event:GS:applicable}
\end{align}
The RVaR crisis event is obviously a special case of
\eqref{eq:crisis:event:GS:applicable}, and the ES crisis event is included as a
limiting case for $v_2 \rightarrow \infty$.  Furthermore, the full conditional
copulas of the underlying joint loss distribution and their inverses are
required to be known as we now explain.  Consider the target density
$\pi= f_{\bX|v_1\leq \bh\T \bX \leq v_2}$.  For its $j$th full conditional
density $\pi_{j|-j}(x_{j}|\bx_{-j})$, notice that
\begin{align*}
  \{ v_1\leq \bh\T \bX \leq v_2, \ \bX_{-j}=\bx_{-j}\} &=
                                                         \left\{\frac{v_1-\tra{\bh_{-j}}\bx_{-j}}{h_j} \leq X_{j}\leq \frac{v_2-\tra{\bh_{-j}}\bx_{-j}}{h_j},\  \bX_{-j}=\bx_{-j}\right\}
\end{align*}
and thus, for $v_{i,j}(\bx_{-j})=(v_i-\tra{\bh_{-j}}\bx_{-j})/h_j$, $i=1,2$, we obtain the cdf of $\pi_{j|-j}$ as
\begin{align}
  F_{X_j|(v_1\leq \bh\T \bX \leq v_2,\ \bX_{-j}=\bx_{-j})}(x_j)
  =\frac{F_{X_{j}|\bX_{-j}=\bx_{-j}}(x_j)-F_{X_{j}|\bX_{-j}=\bx_{-j}}(v_{1,j}(\bx_{-j}))}{F_{X_{j}|\bX_{-j}=\bx_{-j}}(v_{2,j}(\bx_{-j}))-F_{X_{j}|\bX_{-j}=\bx_{-j}}(v_{1,j}(\bx_{-j}))}
  \label{eq:full:conditional:distribution:RVaR}
\end{align}
for $v_{1,j}(\bx_{-j})\leq x_{j} \leq v_{2,j}(\bx_{-j})$.
Denoting the denominator of \eqref{eq:full:conditional:distribution:RVaR} by $\Delta_j(\bx_{-j})$, we obtain the quantile function
\begin{align*}
  F_{X_{j}|(v_1 \leq \bh\T \bX\leq v_2, \ \bX_{-j}=\bx_{-j})}^{-1}(u) = F_{X_{j}|\bX_{-j}=\bx_{-j}}^{-1}\left(\Delta_j(\bx_{-j})\cdot u +F_{X_{j}|\bX_{-j}=\bx_{-j}}(v_{1,j}(\bx_{-j}))\right).
\end{align*}
Therefore, if $F_{X_{j}|\bX_{-j}=\bx_{-j}}$ and its quantile function are
available, one can simulate the full conditional target densities $\pi_{j|-j}$
with the inversion method; see \citet{devroye1985nonuniformgeneration}.
Availability of $F_{X_{j}|\bX_{-j}=\bx_{-j}}$ and its inverse typically depends
on the copula of $\bX$.  By Sklar's theorem $\eqref{eq:sklar:thm}$, the $j$th
full conditional distribution of $F_{\bX}$ can be written as
\begin{align*}
  F_{X_{j}|\bX_{-j}=\bx_{-j}}(x_{j}) = C_{j|-j}(F_j (x_j)\ |\ \bF_{-j}(\bx_{-j})),
\end{align*}
where
$\bF_{(j_1,\dots,j_l)}(\bx_{(j_1,\dots,j_l)})=
(F_{j_1}(x_{j_1}),\dots,F_{j_l}(x_{j_l}))$,
$-(j_1,\dots,j_l)=\{1,\dots,d\}\backslash (j_1,\dots,j_l)$ and $C_{j|-j}$ is the
$j$th \emph{full conditional copula} defined by
\begin{align*}
  C_{j|-j}(u_j|\bu_{-j})=\Prob(U_j \leq u_j \ |\  \bU_{-j}=\bu_{-j})=\frac{D_{-j}C(\bu)}{D_{-j}C(u_1,\dots,u_{j-1},1,u_{j+1},\dots,u_d)},
\end{align*}
where $D$ denotes the operator of partial derivatives with respect to the components given as subscripts and $\bU \sim C$.
Assuming the full conditional copula $C_{j|-j}$ and its inverse $C_{j|-j}^{-1}$ are available, one can simulate $\tilde X_j \sim \pi_{j|-j}$ via
\begin{align*}
  U & \sim \Unif(0,1),\\
  \tilde U &= U + (1-U) C_{j|-j}(F_{j}(v_1(\bx_{-j})\ |\  {\bm F}_{-j}(\bx_{-j})),\\
  \tilde X_{j} &= F_{j}^{-1}\circ C_{j|-j}^{-1} (\tilde U
                 \ |\  {\bm F}_{-j}(\bx_{-j})).
\end{align*}

Examples of copulas for which the full conditional distributions and their
inverses are available include normal, Student $t$, and Clayton
copulas; see \citet{cambou2017quasi}.
In this case
the GS is also applicable to the corresponding survival ($\pi$-rotated) copula
$\hat C$ since
\begin{align*}
  \hat {C}_{j|-j}(\bu) = 1- C_{j|-j}(1-u_j\,|\,\bone_{d'}-\bu_{-j}),\quad
  \hat{C}^{-1}_{j|-j}(\bu) = 1- C_{j|-j}^{-1}(1-u_j\,|\,\bone_{d'}-\bu_{-j}),\quad
  j=1,\dots,d,
\end{align*}
by the relationship $\tilde \bU = \bone-\bU \sim \hat{C}$ for $\bU\sim C$.  In a
similar way, one can also obtain full conditional copulas and their inverses for
other rotated copulas; see
\citet[Section~3.4.1]{hofertkojadinovicmaechleryan2018} for rotated copulas.

In the end, we remark that even if the full conditional distributions and their
inverses are not available, $\pi_{j|-j}$ can be simulated by, for example, the
acceptance and rejection method or even the MH algorithm; see
Appendix~\ref{appendix:subsec:metropolized:gibbs:sampler}.

\subsubsection{Choice of Parameters for GS}\label{subsubsec:choice:parameters:GS}
As discussed in Section~\ref{subsubsec:choice:parameters:HMC}, we use information from the MC presamples to determine the parameters of the Gibbs
kernel~\eqref{eq:GS:proposal}.  Note that standardization of the variables as
applied in the HMC method in Section~\ref{subsubsec:choice:parameters:HMC} is
not available for the GS since the latter changes the underlying joint losss
distribution, and since the copula after rotating variables is in general not
accessible except for in the elliptical case; see \citet{christen2017optimal}.
Among the presented variants of GSs, we adopt RSGS since determining
$d$ probabilities $(\sqc{p}{d})$ is relatively easy whereas RPGS requires $d!$
probabilities to be determined.  To this end, we consider the RSGS with the
parameters $(\sqc{p}{d})$ determined by a heuristic described in
Algorithm~\ref{algorithm:RSGS:heuristic}.
\begin{algorithm}[t]
  \caption{Random scan Gibbs sampler (RSGS) with heuristic to determine $(\sqc{p}{d})$}
  \label{algorithm:RSGS:heuristic}
  \begin{algorithmic}[0]
    \REQUIRE Random number generator of $\pi_{j|-j}$ and $\bx^{(0)} \in \text{supp}(\pi)$.\\
    \INPUT  MC presample $\tilde \bX_{1}^{(0)},\dots,\tilde \bX_{N_0}^{(0)}$, sample size $N\in \IN$, initial state $\bx^{(0)}$, sample size of the pre-run $N_\text{pre}$ and the target autocorrelation $\rho$ ($N_\text{pre}=100$ and $\rho=0.15$ are set as default values).\\
    \OUTPUT $N$ sample path $\bX^{(1)},\dots,\bX^{(N)}$ of the Markov chain.\\ \vspace{1mm}
    \algrule
    \STATE 1) Compute the sample covariance matrix $\hat \Sigma$ based on $\tilde \bX_{1}^{(0)},\dots,\tilde \bX_{N_0}^{(0)}$.\\
    \STATE 2) Set $p_{j} \propto \hat \Sigma_{j,j}-\hat{\Sigma}_{j,-j}\hat{\Sigma}_{-j,-j}^{-1}\hat{\Sigma}_{-j,j}$ and $\bX^{(0)}=\bX_\text{pre}^{(0)}=\bx^{(0)}$.\\
    \STATE 3) \textbf{for} $n:=1,\dots,N_\text{pre}$\\
    \STATE \hspace{8mm} 3-1) Generate $J = j$ with probability $p_j$.\\
    \STATE \hspace{8mm} 3-2) Update $X_{\text{pre},J}^{(n)} \sim \pi_{J|-J}(\cdot| \bX_\text{pre}^{(n-1)})$ and $\bX_{\text{pre},-J}^{(n)}=\bX_{\text{pre},-J}^{(n-1)}$.\\
    \STATE  \hspace{3mm} \textbf{end for}\\
    \STATE 4) Set
    \begin{align*}
      T = \operatorname{argmin}_{h \in \IN_0}\left\{\text{estimated autocorrelations of } \bX_\text{pre}^{(1)},\dots,\bX_\text{pre}^{(N_\text{pre})} \text{ with lag }h\  \leq \rho\right\}.
    \end{align*}\\
    \STATE 5) \textbf{for} $n:=1,\dots,N$, $t:=1,\dots,T$\\
    \STATE \hspace{8mm} 5-1) Generate $J = j$ with probability $p_j$.\\
    \STATE \hspace{8mm} 5-2) Update $X_{J}^{(n-1+t/T)} \sim \pi_{J|-J}(\cdot| \bX^{(n-1+(t-1)/T)})$ and $\bX_{-J}^{(n-1+t/T)}=\bX_{-J}^{(n-1+(t-1)/T)}$.\\
    \STATE  \hspace{3mm} \textbf{end for}\\
  \end{algorithmic}
\end{algorithm}

The RSGS kernel is simulated in Step 3) and 5) of Algorithm~\ref{algorithm:RSGS:heuristic}.
To determine the selection probabilities $\sqc{p}{d}$, consider a one step update of the RSGS from $\bX^{(n)}$ to $\bX^{(n+1)}$ with $\bX^{(n)} \sim \pi$ and the one step kernel
\begin{align*}
  K_\text{RSGS}(\bx,\by)=\sum_{j=1}^d p_j \pi_{j|-j}(y_j|\bx_{-j})\bone_{[\by_{-j}=\bx_{-j}]}.
\end{align*}
\citet[Lemma~3]{liu1995covariance} implies that
\begin{align*}
  \Cov(X_{j}^{(n)},X_{j}^{(n+1)})& = \sum_{i=1}^d p_i \E[\E[X_j\ | \bX_{-i}]]
                                   =  \sum_{i=1}^d p_i \{ m_j^{(2)} - \E[\Var(X_j\ | \ \bX_{-i})]\}
                                   \propto -\sum_{i=1}^d p_i \E[\Var(X_j\ | \ \bX_{-i})]),
\end{align*}
where $m_j^{(k)}$ is the $k$th moment of $\pi_j$.

For the objective function $\sum_{j=1}^d \Cov(X_{j}^{(n)},X_{j}^{(n+1)})$, its
minimizer $(\sqc{p^{\ast}}{d})$ under the constraint $\sum_{j=1}^d p_j =1$
satisfies
\begin{align}
  p_j^{\ast}\propto \E[\Var(X_j\,|\,\bX_{-j})].\label{eq:optimal:selection:probabilities}
\end{align}
While this optimizer can be computed based on the MC presamples, we observed
that its stable estimation is as computationally demanding as estimating the
risk allocations themselves.  Alternatively, we
calculate~\eqref{eq:optimal:selection:probabilities} under the assumption that
$\pi$ follows an elliptical distribution.  Under this assumption,
\eqref{eq:optimal:selection:probabilities} is given by
\begin{align*}
  p_{j} \propto \Sigma_{j,j}-\Sigma_{j,-j}\Sigma_{-j,-j}^{-1}\Sigma_{-j,j}
\end{align*}
where $\Sigma$ is the covariance matrix of $\pi$ and $ \Sigma_{J_1,J_2}$,
$J_1,J_2 \subseteq \{1,\dots,d\}$ is the submatrix of $ \Sigma$ with indices in
$J_1 \times J_2$.  As seen in Step 2) of
Algorithm~\ref{algorithm:RSGS:heuristic}, $\Sigma$ is replaced by its estimate
based on the MC presamples.

As is shown in \citet{christen2017optimal}, Gibbs
samplers 
require a large number of iterations to lower the serial correlation when the
target distribution has strong dependence.  To reduce serial correlations we
take every $T$th sample in Step 5-2), where $T \in \IN$ is called the
\emph{thinning interval of times}.  Note that we use the same notation $T$ as
that of the integration time in HMC since they both represent a repetition time
of some single step.  Based on the preliminary run with length $N_\text{pre}$ in
Step 3) in Algorithm~\ref{algorithm:RSGS:heuristic}, $T$ is determined as the
smallest lag $h$ such that the marginal autocorrelations with lag $h$ are all
smaller than the target autocorrelation $\rho$; see Step 4) in
Algorithm~\ref{algorithm:RSGS:heuristic}.

\section{Numerical Experiments}\label{sec:numerical:experiments}
In this section, we demonstrate the performance of the MCMC methods for
estimating systemic risk allocations by a series of numerical experiments.  We
first conduct a simulation study in which true allocations or their partial
information are available.  Then we perform an empirical study to demonstrate
that our MCMC methods are applicable to a more practical setup. Finally, we
make more detailed comparisons between the MC and MCMC methods in various
setups.  All experiments are run on a MacBook Air with 1.4 GHz Intel Core i5
processor and 4 GB 1600 MHz of DDR3 RAM.

\subsection{Simulation Study}\label{subsec:simulation:study}
In this simulation study, we compare the estimates and standard errors of the MC
and MCMC methods under the low-dimensional risk models described in
Section~\ref{subsubsec:model:description}.  The results and discussions are
summarized in Section~\ref{subsubsec:results:discussions}.

\subsubsection{Model Description}\label{subsubsec:model:description}
We consider the following three-dimensional loss distributions:

\begin{enumerate}
\item \emph{generalized Pareto distributions (GPDs)} with parameters
  $(\xi_j,\beta_j)=(0.3,1)$ and survival Clayton copula with parameter
  $\theta=2$ so that Kendall's tau equals $\tau = \theta/(\theta+2)=0.5$;
\item multivariate Student $t$ distribution with $\nu = 5$ degrees of freedom,
  location vector $\bzero$ and dispersion matrix $\Sigma = (\rho_{i,j})$ where
  $\rho_{j,j}=1$ and $\rho_{i,j}=|i-j|/d$ for $i,j=1,\dots,d$, $i\neq j$.
\end{enumerate}

Since the marginals are homogeneous and the copula is exchangeable, the
systemic risk allocations under the loss distribution (M1) are all equal provided that the
crisis event is invariant under the permutation of the variables.  For the loss distribution (M2), by
ellipticality of the joint distribution, analytical formulas of risk
contribution type systemic risk allocations are available; see
\citet[Corollary 8.43]{mcneil2015quantitative}.  The parameters of the distributions (M1)
and (M2) take into account the stylized facts that the loss distribution is
heavy-tailed and extreme losses are positively dependent.

We consider the VaR, RVaR and ES crisis events with confidence levels
$\alpha^{\operatorname{VaR}}=0.99$, $(\alpha_1^{\RVaR},\alpha_2^{\RVaR})=(0.975,0.99)$
and $\alpha^{\ES}=0.99$, respectively.  For each crisis event, the risk
contribution, VaR, RVaR and ES type systemic risk allocations are estimated by
the MC and MCMC methods, where the parameters of the marginal risk measures VaR,
RVaR and ES are set to be $\beta^{\operatorname{VaR}}=0.99$,
$(\beta_1^{\RVaR},\beta_2^{\RVaR})=(0.975,0.99)$ and $\beta^{\ES}=0.99$,
respectively.

We first conduct the MC simulation for the distributions (M1) and (M2).  For
the VaR crisis event, the modified event
$\mathcal C^{\text{mod}} = \{\VaR_{\alpha-\delta}(S) \leq \bone_d\T \bx \leq
\VaR_{\alpha+\delta}(S)\}$ with $\delta = 0.001$ is used to ensure that
$\Prob(\bX \in \mathcal C^{\text{mod}})>0$.  Based on these MC presamples, the Markov chains
are constructed as described in Sections~\ref{subsec:HMC}
and~\ref{subsec:gibbs:samplers}.  For the MCMC method, (M1) is the case of pure
losses and (M2) is the case of P$\&$L.  Therefore, the HMC method is applied to the distribution (M1)
for the VaR and RVaR crisis events, the GS is applied to (M1) for the ES crisis
event and the GS is applied to the distribution (M2) for the RVaR and ES crisis events.  The
target distribution of (M2) with VaR constraint is free from constraints and was
already investigated in \citet{koike2019estimation}; we thus omit this case and
consider the five remaining cases.

Note that 99.8\% of the MC samples from the unconditional distribution are
discarded for the VaR crisis event and a further 97.5\% of them are wasted to
estimate the RVaR contributions.  Therefore, $1/(0.002 \times 0.025)= 10^5 / 5 =20,000$ MC samples are required to obtain
one MC sample from the conditional distribution.  Taking this into
account, the sample size of the MC estimator is set to be $N_\text{MC}=10^5$.
The sample size of the MCMC estimators is free from such constraints and thus is
chosen to be $N_\text{MCMC}=10^4$.  Initial values $\bx_0$ for the MCMC methods are
taken as the mean vector calculated from the MC samples.  Biases are computed only
for the contribution type allocations in the distribution (M2) since the true values are
available in this case.  For all the five cases, the MC and the MCMC
standard errors are computed according to \citet[Chapter 1]{glasserman2013monte}
for MC, and \citet{jones2006fixed} for MCMC.  Asymptotic variances of the MCMC
estimators are estimated by the batch means estimator with batch length
$L_{N}:=\lceil N^{\frac{1}{2}} \rceil = 100$ and batch size
$B_{N}:=\lceil N/L_{N} \rceil = 100$.  The results are summarized in
Tables~\ref{Table:HMC} and~\ref{Table:GS}.
\begin{table}[htbp]
  \centering
  \begin{tabular}{
    l
    r
    r
    r
    r
    r
    r
    }
    \toprule
     & \multicolumn{3}{c}{MC} & \multicolumn{3}{c}{HMC}\\
    \cmidrule(lr{0.4em}){2-4} \cmidrule(lr{0.4em}){5-7}
    \multicolumn{1}{l}{Estimator} & \multicolumn{1}{c}{$A_1^{\varrho,\mathcal C}(\bX)$} & \multicolumn{1}{c}{$A_2^{\varrho,\mathcal C}(\bX)$} & \multicolumn{1}{c}{$A_3^{\varrho,\mathcal C}(\bX)$} & \multicolumn{1}{c}{$A_1^{\varrho,\mathcal C}(\bX)$} & \multicolumn{1}{c}{$A_2^{\varrho,\mathcal C}(\bX)$} & \multicolumn{1}{c}{$A_3^{\varrho,\mathcal C}(\bX)$}\\
    \midrule
    \multicolumn{7}{l}{{(I) GPD + survival Clayton with VaR crisis event}: $\{S = \VaR_{0.99}(S)\}$}\\[4pt]
    $\E[\bX|\mathcal C^{\operatorname{VaR}}]$ & 9.581 & 9.400 & 9.829  & 9.593 & 9.599 & 9.619\\
    \textbf{Standard error} &   0.126 & 0.118 & 0.120  & 0.007 & 0.009 & 0.009\\  [4pt]
    $\RVaR_{0.975,0.99}(\bX|\mathcal C^{\operatorname{VaR}})$ &12.986 & 12.919 & 13.630  & 13.298 & 13.204 & 13.338\\
    \textbf{Standard error} &0.229 & 0.131 & 0.086 & 0.061 & 0.049 & 0.060 \\  [4pt]
    $\operatorname{VaR}_{0.99}(\bX|\mathcal C^{\operatorname{VaR}})$ & 13.592 & 13.235 & 13.796  & 13.742 & 13.565 & 13.768 \\
    \textbf{Standard error} &0.647 & 0.333 & 0.270  & 0.088 & 0.070 & 0.070 \\ [4pt]
    $\ES_{0.99}(\bX|\mathcal C^{\operatorname{VaR}})$ & 14.775 & 13.955 & 14.568 & 14.461 & 14.227 & 14.427\\
    \textbf{Standard error} &  0.660 & 0.498 & 0.605  & 0.192 & 0.176 & 0.172 \\ [8pt]
    \multicolumn{7}{l}{(II) GPD + Survival Clayton with RVaR crisis event: $\{\VaR_{0.975}(S)\leq S \leq \VaR_{0.99}(S)\}$ }\\[4pt]
    $\E[\bX|\mathcal C^{\operatorname{RVaR}}]$ & 7.873 & 7.780 & 7.816  & 7.812 & 7.802 & 7.780\\
    \textbf{Standard error} &0.046 & 0.046 & 0.046 &  0.012 & 0.012 & 0.011\\   [4pt]
    $\RVaR_{0.975,0.99}(\bX|\mathcal C^{\operatorname{RVaR}})$ &11.790 & 11.908 & 11.680 & 11.686 & 11.696 & 11.646 \\
    \textbf{Standard error} & 0.047 & 0.057 & 0.043  & 0.053 & 0.055 & 0.058\\  [4pt]
    $\operatorname{RVaR}_{0.99}(\bX|\mathcal C^{\operatorname{VaR}})$ & 12.207 & 12.382 & 12.087 &  12.102 & 12.053 & 12.044\\
    \textbf{Standard error} & 0.183 & 0.197 & 0.182 & 0.074 & 0.069 & 0.069 \\  [4pt]
    $\ES_{0.99}(\bX|\mathcal C^{\operatorname{RVaR}})$ & 13.079 & 13.102 & 13.059 &  12.859 & 12.791 & 12.713\\
    \textbf{Standard error} &0.182 & 0.173 & 0.188  & 0.231 & 0.218 & 0.187\\
    \bottomrule
  \end{tabular}
  \caption{Estimates and standard errors for the MC and HMC estimators of risk contribution, RVaR, VaR and ES type systemic risk allocations under (I) the VaR crisis event and (II) the RVaR crisis event for the loss distribution (M1). The sample size of the MC method is $N_\text{MC}=10^5$ and that of the HMC method is $N_\text{MCMC}=10^4$. The acceptance rate (ACR), stepsize $\epsilon$, integration time $T$ and run time are 
  ACR $=0.996$,  $\epsilon=0.210$, $T = 12$ and run time $= 1.277$ mins in Case (I), and ACR $=0.984$,  $\epsilon=0.095$, $T = 13$ and run time $= 1.649$ mins in Case (II).
  }
    \label{Table:HMC}
\end{table}

\begin{table}[htbp]
  \centering
    \begin{tabular}{
    l
    rrrrrr
    }
    \toprule
     & \multicolumn{3}{c}{MC} & \multicolumn{3}{c}{GS}\\
    \cmidrule(lr{0.4em}){2-4} \cmidrule(lr{0.4em}){5-7}
    \multicolumn{1}{l}{Estimator} & \multicolumn{1}{c}{$A_1^{\varrho,\mathcal C}(\bX)$} & \multicolumn{1}{c}{$A_2^{\varrho,\mathcal C}(\bX)$} & \multicolumn{1}{c}{$A_3^{\varrho,\mathcal C}(\bX)$} & \multicolumn{1}{c}{$A_1^{\varrho,\mathcal C}(\bX)$} & \multicolumn{1}{c}{$A_2^{\varrho,\mathcal C}(\bX)$} & \multicolumn{1}{c}{$A_3^{\varrho,\mathcal C}(\bX)$}\\
    \midrule
    \multicolumn{7}{l}{{(III) GPD + survival Clayton with ES crisis event}: $\{\VaR_{0.99}(S)\leq S\}$}\\[4pt]
      $\E[\bX|\mathcal C^{\operatorname{ES}}]$ &15.657 & 15.806 & 15.721  & 15.209 & 15.175 & 15.190\\
      \textbf{Standard error} &0.434 & 0.475 & 0.395 &  0.257 & 0.258 & 0.261 \\  [4pt]
      $\RVaR_{0.975,0.99}(\bX|\mathcal C^{\operatorname{ES}})$ &41.626 & 41.026 & 45.939  & 45.506 & 45.008 & 45.253\\
      \textbf{Standard error} & 1.211 & 1.065 & 1.615 &  1.031 & 1.133 & 1.256\\  [4pt]
      $\operatorname{VaR}_{0.99}(\bX|\mathcal C^{\operatorname{ES}})$ & 49.689 & 48.818 & 57.488  & 55.033 & 54.746 & 54.783\\
      \textbf{Standard error} &4.901 & 4.388 & 4.973  & 8.079 & 5.630 & 3.803\\ [4pt]
      $\ES_{0.99}(\bX|\mathcal C^{\operatorname{ES}})$ & 104.761 & 109.835 & 97.944 & 71.874 & 72.588 & 70.420\\
      \textbf{Standard error} &23.005 & 27.895 & 17.908  & 4.832 & 4.584 & 4.313\\ [8pt]
      \multicolumn{7}{l}{(IV) Multivariate $t$ with RVaR crisis event: $\{\VaR_{0.975}(S)\leq S \leq \VaR_{0.99}(S)\}$ }\\[4pt]
      $\E[\bX|\mathcal C^{\operatorname{RVaR}}]$ & 2.456 & 1.934 & 2.476 & 2.394 & 2.060 & 2.435\\
      \textbf{Bias} & 0.019 & -0.097 & 0.038 &  -0.043 & 0.029 & -0.002\\
      \textbf{Standard error} & 0.026 & 0.036 & 0.027 & 0.014 & 0.023 & 0.019\\   [4pt]
      $\RVaR_{0.975,0.99}(\bX|\mathcal C^{\operatorname{RVaR}})$ &  4.670 & 4.998 & 4.893  & 4.602 & 5.188 & 4.748\\
      \textbf{Standard error} & 0.037 & 0.042 & 0.031  & 0.032 & 0.070 & 0.048\\  [4pt]
      $\operatorname{RVaR}_{0.99}(\bX|\mathcal C^{\operatorname{VaR}})$ &  5.217 & 5.397 & 5.240 &  4.878 & 5.717 & 5.092\\
      \textbf{Standard error} & 0.238 & 0.157 & 0.145  & 0.049 & 0.174 & 0.100\\  [4pt]
      $\ES_{0.99}(\bX|\mathcal C^{\operatorname{RVaR}})$ & 5.929 & 5.977 & 5.946 &  5.446 & 6.517 & 6.063\\
      \textbf{Standard error} &  0.204 & 0.179 & 0.199 &  0.156 & 0.248 & 0.344\\  [8pt]
      \multicolumn{7}{l}{(V) Multivariate $t$ with ES crisis event: $\{\VaR_{0.99}(S)\leq S\}$}\\[4pt]
      $\E[\bX|\mathcal C^{\operatorname{ES}}]$ & 3.758 & 3.099 & 3.770 &  3.735 & 3.126 & 3.738\\
      \textbf{Bias} & 0.017 & -0.018 & 0.029  & -0.005 & 0.009 & -0.003 \\
      \textbf{Standard error} &0.055 & 0.072 & 0.060  & 0.031 & 0.027 & 0.030\\   [4pt]
      $\RVaR_{0.975,0.99}(\bX|\mathcal C^{\operatorname{ES}})$ & 8.516 & 8.489 & 9.051  & 8.586 & 8.317 & 8.739\\
      \textbf{Standard error} & 0.089 & 0.167 & 0.161  & 0.144 & 0.156 & 0.158\\  [4pt]
      $\operatorname{VaR}_{0.99}(\bX|\mathcal C^{\operatorname{ES}})$ &  9.256 & 9.754 & 10.327 &  9.454 & 9.517 & 9.890 \\
      \textbf{Standard error} & 0.517 & 0.680 & 0.698 & 0.248 & 0.293 & 0.327 \\  [4pt]
      $\ES_{0.99}(\bX|\mathcal C^{\operatorname{ES}})$ & 11.129 & 12.520 & 12.946  & 11.857 & 12.469 & 12.375\\
      \textbf{Standard error} & 0.595 & 1.321 & 0.826 &  0.785 & 0.948 & 0.835 \\
      \bottomrule
    \end{tabular}
    \caption{Estimates and standard errors for the MC and the GS estimators
      of risk contribution, VaR, RVaR and ES type systemic risk allocations under
      (III) distribution (M1) and the ES crisis event, (IV) distribution (M2) and the RVaR crisis event, and
      (V) distribution (M2) and ES crisis event.  The sample size of the MC method is
      $N_\text{MC}=10^5$ and that of the GS is $N_\text{MCMC}=10^4$.  The thinning
      interval of times $T$, selection probability $\bp$ and run time are
      $T=12$, $\bp = (0.221, 0.362, 0.416)$ and run time $=107.880$ secs in Case (III),
      $T=10$, $\bp = (0.330, 0.348, 0.321)$ and run time $=56.982$ secs in Case (IV) and
      $T=4$, $\bp = (0.241, 0.503,  0.255)$ and run time $=22.408$ secs in Case (V).
      }
    \label{Table:GS}
\end{table}

\subsubsection{Results and Discussions}\label{subsubsec:results:discussions}
Since fast random number generators are available for the joint loss
distributions (M1) and (M2), the MC estimators are computed almost
instantly.  On the other hand, the MCMC methods cost around 1.5 minutes for
simulating the $N=10^4$ MCMC samples as reported in Tables~\ref{Table:HMC}
and~\ref{Table:GS}.  For the HMC method, the main computational cost consists of
calculating gradients $N\times T$ times for the leapfrog method, and calculating
the ratio of target densities $N$ times in the acceptance/rejection step, where
$N$ is the length of the sample path and $T$ is the integration time.  For the
GS, simulating an $N$-sample path requires $N\times T\times d$ random numbers
from the full conditional distributions where $T$ here is the thinning interval
of times.  Therefore, the computational time of the GS linearly increases
w.r.t.\ the dimension $d$, which can become prohibitive for the GS in high
dimensions.  To save computational time, MCMC methods in general require careful
implementations of calculating the gradients and the ratio of the target densities for HMC, and
of simulating the full conditional distributions for GS.

\begin{figure}[H]
  \centering
  \vspace{-30mm}
  \includegraphics[width=15 cm]{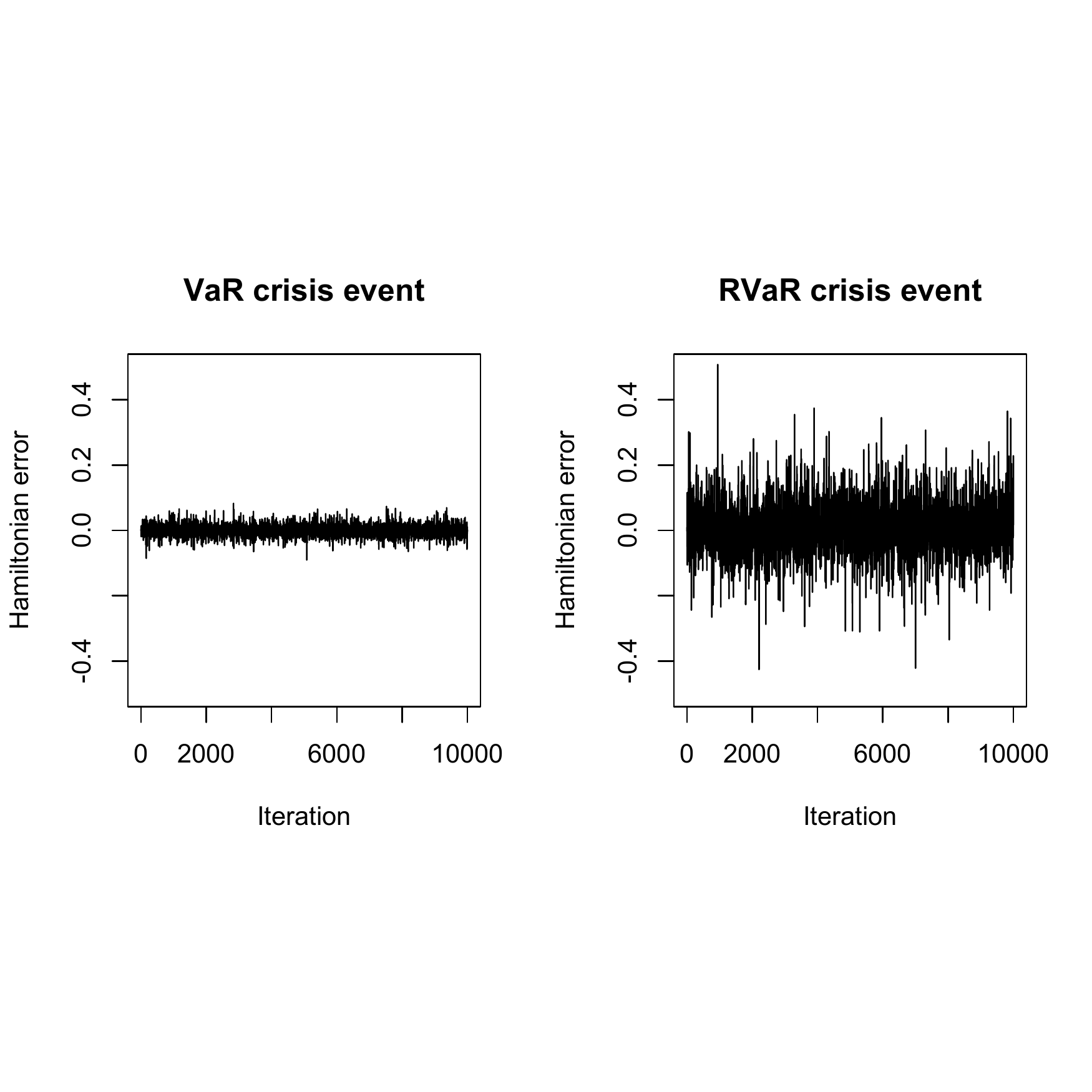}\vspace{-35mm}
  \caption{Hamiltonian errors of the HMC methods for estimating systemic risk allocations with VaR (left) and RVaR (right) crisis events for the loss distribution (M1).
    The stepsize and the integration time are set to be $(\epsilon,T)=(0.210,12)$ in Case (I) and $(\epsilon,T)=(0.095,13)$ in Case (II).
}
  \label{fig:hamiltonian:error}
\end{figure}

Next, we inspect the performance of the HMC and GS methods.  We observed
  that autocorrelations of all sample paths steadily decreased below 0.1 if lags
  are larger than 15. Together with the high ACRs, we conclude that the Markov
  chains can be considered to be converged.  According to the heuristic in
Algorithm~\ref{algorithm:adaptive:stepsize:integratio:time}, the stepsize and
the integration time for the HMC method are selected to be $(\epsilon,T)=(0.210,12)$ in Case (I) and $(\epsilon,T)=(0.095,13)$ in Case (II).
 As indicated by the
small Hamiltonian errors in Figure~\ref{fig:hamiltonian:error}, the acceptance
rates in both cases are quite close to 1.

For the GS, the thinning interval of times $T$ and the selection probability
$\bp$ are determined as $T=12$ and $\bp = (0.221, 0.362, 0.416)$ in Case (III), $T=10$ and $\bp = (0.330, 0.348, 0.321)$ in Case (IV) and $T=4$ and $\bp = (0.241, 0.503,  0.255)$ in Case (V).
For biases of the estimators, observe that in all cases ((I) to (V)), the
estimates of the MC method and the MCMC method are close to each other.
In Cases (I), (II) and (III), the true allocations are the homogeneous allocations whereas their exact values are not known.  From the estimates in
Tables~\ref{Table:HMC} and~\ref{Table:GS}, the MCMC estimates are on average more equally
allocated compared to those of the MC method especially in Case 
(III) where heavy-tailedness may lead to quite slow convergence rates of the MC
method.  Therefore, lower biases of the MCMC estimators are obtained compared
to those of the MC estimators.  In the case of risk contributions in Case 
(IV) and (V), exact biases are computed based on ellipticality, and they show
that the GS estimator has a smaller bias than the one of the MC estimator.

Although the MC sample size is 10 times larger than that of the MCMC method, the
standard error of the latter is in most cases smaller than the MC standard
error.  This improvement becomes larger as the probability of the crisis event
becomes smaller.  The largest improvement is observed in Case (I) with VaR
crisis event and the smallest one is in Cases (III) and (V) with ES crisis
event.  MCMC estimates of the risk contribution type allocations have
consistently smaller standard errors than the MC ones.  For the RVaR, VaR and
ES type allocations, the improvement of standard error varies according to the loss models
and the crisis event.  A notable improvement is observed for ES type allocation
in Case (III) although a stable statistical inference is challenging due to the
heavy-tailedness of the target distribution.

Overall, the simulation study shows that the MCMC estimators outperform the
MC estimators due to the increased effective sample size and its
insusceptibility to the probability of the crisis event.  The MCMC estimators
are especially recommended when the probability of the crisis event is too small
for the MC method to simulate sufficiently many samples for a meaningful
statistical analysis.
\begin{Remark}[Joint loss distributions with negative dependence in the tail]
  In the above simulation study, we only considered joint loss distributions
  with positive dependence.  Under the existence of positive dependence, the
  target density $f_{\bX|v_{\alpha} \leq S\leq v_{\beta}}$ puts more probability
  mass around its mean, and the probability decays as the 
  point moves away from the mean since positive dependence among $X_1,\dots,X_d$
  prevents them from going in opposite directions (i.e., one component increases
  and another one decreases) under the sum constraint; see
  \citet{koike2019estimation} for details.  This phenomenon leads to the target
  distributions more centered and elliptical, which in turn
  facilitates efficient moves of Markov chains.  Although it may not be
  realistic, joint loss distributions with negative dependence in the tail are
  also possible.  In this case, the target distribution has more variance, heavy
  tails and is even multimodal since two components can move in opposite
  directions under the sum constraint.  For such cases, constructing efficient
  MCMC methods becomes more challenging; see \citet{lan2014wormhole} for a
  remedy for multimodal target distributions with Riemannian manifold HMC.
\end{Remark}

\subsection{Empirical Study}\label{subsec:empirical:study}
\begin{figure}[htbp]
  \centering
  \includegraphics[width=15 cm]{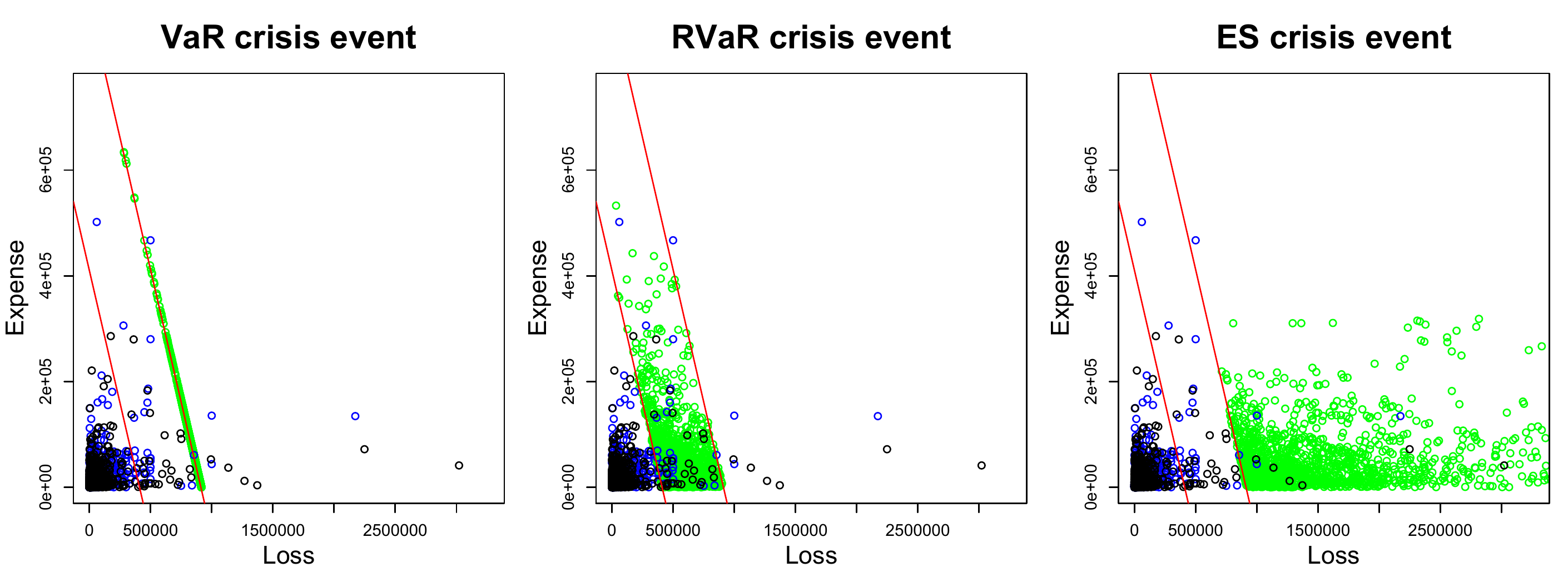}
  \caption{Plots of $N = 1,500$ MCMC samples (green) with VaR (left), RVaR
    (center) and ES (right) crisis events.  All plots include the data and the
    MC samples with sample size $N = 1,500$ in black and blue dots,
    respectively.  The red lines represent
    $x_1 + x_2 = \widehat{\VaR}_{\alpha_1}(S)$ and
    $x_1 + x_2 = \widehat{\VaR}_{\alpha_2}(S)$ where
    $\widehat{\VaR}_{\alpha_1}(S)=4.102\times 10^4$ and
    $\widehat{\VaR}_{\alpha_2}(S)=9.117 \times 10^4$ are the MC
    estimates of $\VaR_{\alpha_1}(S)$ and $\VaR_{\alpha_2}(S)$, respectively, for
    $\alpha_1=0.975$ and $\alpha_2 = 0.99$.}
  \label{fig:data:MCMC:plot}
\end{figure}

In this section, we illustrate our suggested MCMC methods for estimating risk
allocations from insurance company indemnity claims.  The dataset consists of 1,500
liability claims provided by Insurance Services Office.  Each claim contains an
indemnity payment $X_1$ and an allocated loss adjustment expense (ALAE) $X_2$;
see \citet{hogg2009loss} for a description.  The joint distribution of losses
and expenses is studied, for example in \citet{frees1998understanding} and
\citet{klugman1999fitting}.  Based on \citet{frees1998understanding}, we adopt
the following parametric model:

\begin{enumerate}
\item[(M3)] univariate marginals are
  $X_1\sim \Par(\lambda_1,\theta_1)$ and
  $X_2\sim \Par(\lambda_2,\theta_2)$ with
  $(\lambda_1,\theta_1) = (14,036, 1.122)$ and
  $(\lambda_2,\theta_2)= (14,219, 2.118)$, and the copula is the survival
  Clayton copula with parameter $\theta = 0.512$ (which corresponds to
  Spearman's rho $\rho_{\text{S}}=0.310$).
\end{enumerate}
Note that in the loss distribution (M3) the Gumbel copula used in
  \citet{frees1998understanding} is replaced by the survival Clayton copula
  since both of them have the same type of tail dependence and the latter
  possesses more computationally tractable derivatives. The parameter of the
  survival Clayton copula is determined so that it reaches the same Spearman's
  rho observed in \citet{frees1998understanding}.
Figure~\ref{fig:data:MCMC:plot} illustrates the data and samples from the distribution
(M3).  Our goal is to calculate the VaR, RVaR and ES type allocations with VaR,
RVaR and ES crisis events for the same confidence levels as in
Section~\ref{subsubsec:model:description}.  We apply the HMC method to all three
crisis events since, due to the infinite and finite variances of $X_1$ and
$X_2$, respectively, the optimal selection probability of the second variable
calculated in Step 2) of Algorithm~\ref{algorithm:RSGS:heuristic} is quite close
to 0, and thus the GS did not perfrom well.  The simulated HMC samples are
illustrated in Figure~\ref{fig:data:MCMC:plot}.  The results of estimating the
systemic risk allocations are summarized in Table~\ref{Table:HMC:emp}.

\begin{figure}[htbp]
  \centering
  \vspace{-40mm}
  \includegraphics[width=15 cm]{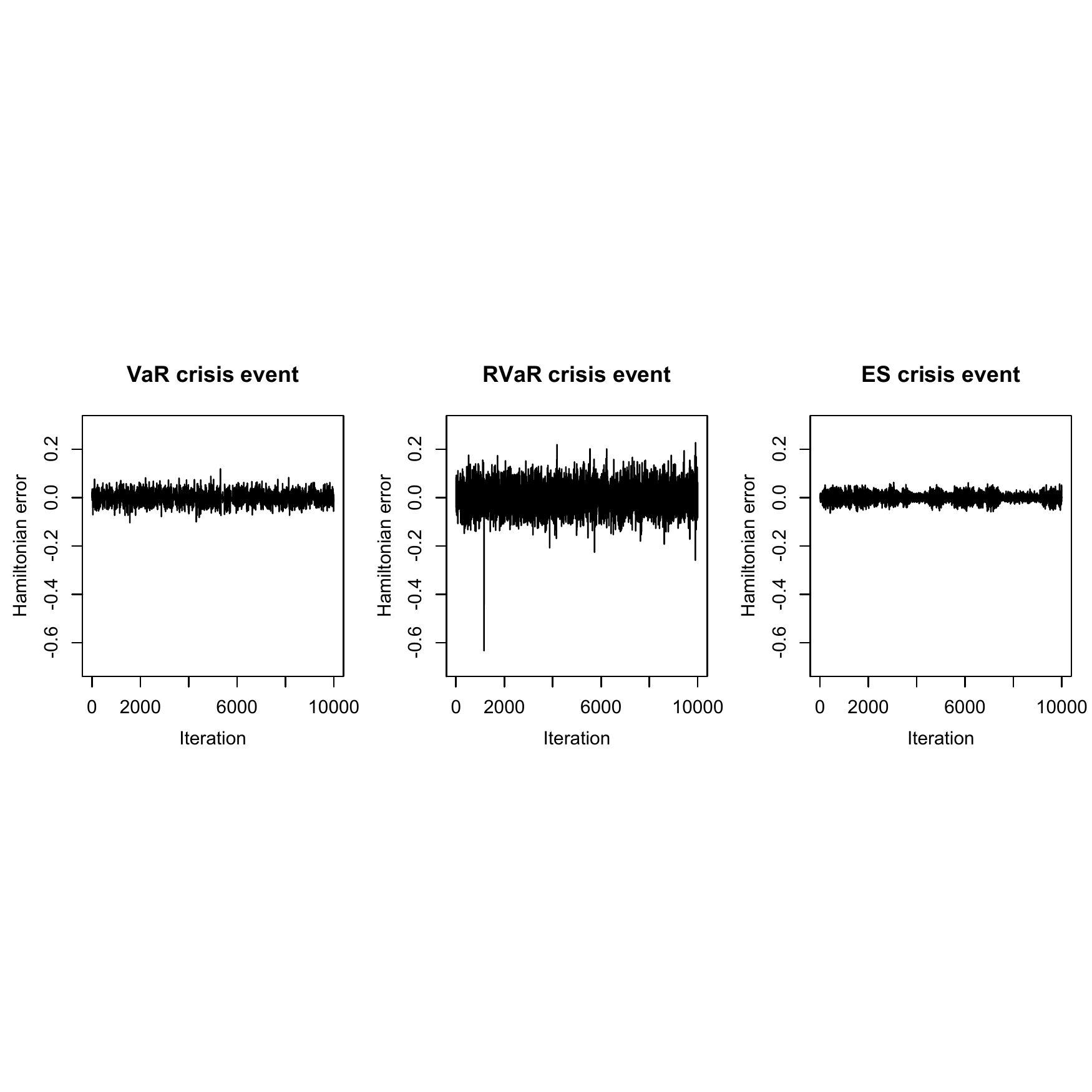}
  \vspace{-45mm}
  \caption{Hamiltonian errors of the HMC methods for estimating systemic risk allocations with VaR, RVaR and ES crisis events for the loss distribution (M3).
    The stepsize and the integration time are chosen as 
    $(\epsilon,T)=(0.015,34)$, $(\epsilon,T)=(0.026,39)$ and $(\epsilon,T)=(5.132\times 10^{-5},838)$, respectively.
  }
  \label{fig:hamiltonian:error:emp}
\end{figure}

The HMC samples shown in Figure~\ref{fig:data:MCMC:plot} indicate that the
conditional distributions of interest are successfully simulated from the
desired regions.  As displayed in Figure~\ref{fig:hamiltonian:error:emp}, the
Hamiltonian errors of all three HMC methods are sufficiently small, which leads
to the high ACRs of $0.997$, $0.986$ and $0.995$ as listed in Table~\ref{Table:HMC:emp}.
We also observed that autocorrelations of all sample paths steadily
  decreased below 0.1 if lags are larger than 80. Together with the high ACRs,
  we conclude that the Markov chains can be considered to be converged.  Due to the
heavy-tailedness of the target distribution in the case of the ES crisis event, the
stepsize is very small and the integration time is very large compared to the
former two cases of the VaR and RVaR crisis events.  As a result, the HMC
algorithm in this case has a long run time.

The estimates of the MC and HMC methods are close in all cases except Case 
(III).  In Case (III), the HMC estimates are smaller than the MC ones in almost
all cases.  Based on the much smaller standard errors of HMC, one could
  infer that the MC estimates are likely overestimating the allocations due to a
  small number of extremely large losses although the corresponding conditional
  distribution is extremely heavy-tailed and thus no estimation method might be
  reliable.  In terms of the standard error, the estimation of systemic risk allocations by
the HMC method is improved in Cases (I) and (III) compared to that of the MC
method; the MC standard errors are slightly smaller than those of HMC in Case 
(II).  All results considered, we conclude from this empirical study that the
MCMC estimators outperform the MC estimators in terms of standard error.  On the
other hand, as indicated by the theory of HMC with normal kinetic energy, the
HMC method is not recommended for heavy-tailed target distributions due to a
long computational time caused by a small stepsize and large integration time determined by Algorithm~\ref{algorithm:RSGS:heuristic}.

\begin{table}[htbp]
  \centering
    \begin{tabular}{
    l
    rrrr
    }
    \toprule
      & \multicolumn{2}{c}{MC} & \multicolumn{2}{c}{HMC}\\
      \cmidrule(lr{0.4em}){2-3} \cmidrule(lr{0.4em}){4-5}
      \multicolumn{1}{l}{Estimator} & \multicolumn{1}{c}{$A_1^{\varrho,\mathcal C}(\bX)$} & \multicolumn{1}{c}{$A_2^{\varrho,\mathcal C}(\bX)$} & \multicolumn{1}{c}{$A_1^{\varrho,\mathcal C}(\bX)$} & \multicolumn{1}{c}{$A_2^{\varrho,\mathcal C}(\bX)$}\\
      \midrule
      \multicolumn{5}{l}{(I) VaR crisis event: $\{S = \VaR_{0.99}(S)\}$}\\[4pt]
      $\E[\bX|\mathcal C^{\operatorname{VaR}}]$ &842465.497 & 73553.738  & 844819.901 & 71199.334\\
      \textbf{Standard error} &   7994.573 & 7254.567 &  6306.836 & 6306.836\\  [4pt]
      $\RVaR_{0.975,0.99}(\bX|\mathcal C^{\operatorname{VaR}})$ &  989245.360 & 443181.466 & 915098.833 & 428249.307\\
      \textbf{Standard error} &307.858 & 24105.163 &  72.568 & 20482.914 \\  [4pt]
      $\operatorname{VaR}_{0.99}(\bX|\mathcal C^{\operatorname{VaR}})$ & 989765.514 & 500663.072  & 915534.362 & 615801.118\\
      \textbf{Standard error} & 4670.966 & 54576.957 &  669.853 & 96600.963 \\ [4pt]
      $\ES_{0.99}(\bX|\mathcal C^{\operatorname{VaR}})$ & 990839.359 & 590093.887 & 915767.076 & 761038.843\\
      \textbf{Standard error} & 679.055 & 75024.692 & 47.744 & 31211.908 \\ [8pt]
      \multicolumn{5}{l}{(II) RVaR crisis event: $\{\VaR_{0.975}(S)\leq S \leq \VaR_{0.99}(S)\}$ }\\[4pt]
      $\E[\bX|\mathcal C^{\operatorname{RVaR}}]$ & 528455.729 & 60441.368 & 527612.751 & 60211.561\\
      \textbf{Standard error} & 3978.477 & 2119.461  & 4032.475 & 2995.992\\   [4pt]
      $\RVaR_{0.975,0.99}(\bX|\mathcal C^{\operatorname{RVaR}})$ & 846956.570 & 349871.745 & 854461.670 & 370931.946\\
      \textbf{Standard error} & 1866.133 & 6285.523 &  2570.997 & 9766.697\\  [4pt]
      $\operatorname{VaR}_{0.99}(\bX|\mathcal C^{\operatorname{RVaR}})$ & 865603.369 & 413767.829  & 871533.550 & 437344.509\\
      \textbf{Standard error} &  5995.341 & 29105.059 &  12780.741 & 21142.135 \\  [4pt]
      $\ES_{0.99}(\bX|\mathcal C^{\operatorname{RVaR}})$ & 882464.968 & 504962.099  & 885406.811 & 529034.580\\
      \textbf{Standard error} & 3061.110 & 17346.207 &  3134.144 & 23617.278\\ [8pt]
      \multicolumn{5}{l}{(III) ES crisis event: $\{\VaR_{0.99}(S)\leq S\}$ }\\[4pt]
      $\E[\bX|\mathcal C^{\operatorname{ES}}]$ & 8663863.925 & 137671.653  & 2934205.458 & 140035.782\\
      \textbf{Standard error} &  3265049.590 & 10120.557 &  165794.772 & 14601.958\\  [4pt]
      $\RVaR_{0.975,0.99}(\bX|\mathcal C^{\operatorname{ES}})$ &  35238914.131 & 907669.462  & 17432351.450 & 589309.196\\
      \textbf{Standard error} & 2892208.689 & 31983.660 &  443288.649 & 3471.641 \\  [4pt]
      $\operatorname{VaR}_{0.99}(\bX|\mathcal C^{\operatorname{ES}})$ & 56612082.905 & 1131248.055 & 20578728.307 & 615572.940\\
      \textbf{Standard error} & 1353975.612 & 119460.411 &  1364899.752 & 12691.776 \\ [4pt]
      $\ES_{0.99}(\bX|\mathcal C^{\operatorname{ES}})$ & 503537848.192 & 2331984.181  & 25393466.446 & 649486.810\\
      \textbf{Standard error} &   268007317.199 & 468491.127  & 1138243.137 & 7497.200 \\
      \bottomrule
    \end{tabular}
    \caption{Estimates and standard errors for the MC and HMC estimators of RVaR,
      VaR and ES type systemic risk allocations under the loss distributon (M3) with the (I)
      VaR crisis event, (II) RVaR crisis event and (III) ES crisis event.  The
      MC sample size is $N_\text{MC}=10^5$ and that of the HMC method is
      $N_\text{MCMC}=10^4$.  The acceptance rate (ACR), stepsize $\epsilon$,
      integration time $T$ and run time are 
      ACR $=0.997$, $\epsilon=0.015$, $T=34$ and run time $=2.007$ mins in Case (I),
      ACR $=0.986$, $\epsilon=0.026$, $T=39$ and run time $=2.689$ mins in Case (II),
      ACR $=0.995$, $\epsilon=5.132\times 10^{-5}$, $T=838$ and run time $=44.831$ mins in Case (III).
    }
    \label{Table:HMC:emp}
\end{table}

  \subsection{Detailed Comparison of MCMC with MC}\label{subsec:detailed:comparison:MCMC:MC}
  In the previous numerical experiments, we fixed the dimensions of the
  portfolios and confidence levels of the crisis events. Comparing the MC and
  MCMC methods after balancing against computational time might be more
  reasonable although one should keep in mind that run time depends on various
  external factors, such as the implementation, hardware, workload, programming
  language or compiler options (and our implementation was not optimized for any
  of these factors). In this section, we compare the MC and MCMC methods with
  different dimensions, confidence levels, and parameters of the HMC methods in
  terms of bias, standard error and the mean squared error (MSE), adjusted
  by run time.

 In this experiment, we fix the sample size of the MC and MCMC methods
  as $N_{\text{MC}}=N_{\text{MCMC}}=10^4$.  In addition, we assume
  $\bX \sim t_{\nu}(\bzero, P)$, i.e., the joint loss follows the multivariate
  Student $t$ distribution with $\nu=6$ degrees of freedom, location vector
  $\bzero$ and dispersion matrix $P$, which is the correlation matrix with all
  off-diagonal entries equal to $1/12$. The dimension $d$ of the loss
  portfolio will vary for comparison.  We consider only risk contribution type
  systemic risk allocations under VaR, RVaR and ES crisis events as true values
  of these allocations are available to compare against; see
  \citet[Corollary~8.43]{mcneil2015quantitative}. If $b$ and $\sigma$ denote the
  bias and standard deviation of the MC or MCMC estimator and $S$ the run time, then
  (under the assumption that run time linearly increases by
  sample size) we define the \emph{time-adjusted MSEs} by
  \begin{align*}
    \text{MSE}_{\text{MC}}=b_{\text{MC}}^2 + \frac{\sigma_{\text{MC}}^2}{\frac{S_{\text{MCMC}}}{S_{\text{MC}}} \times N_{\text{MCMC}}}\quad \text{and}\quad
    \text{MSE}_{\text{MCMC}}=b_{\text{MCMC}}^2 + \frac{\sigma_{\text{MCMC}}^2}{N_{\text{MCMC}}}.
  \end{align*}

 We can then compare the MC and MCMC estimators in terms of bias,
  standard error and time-adjusted MSE under the following three scenarios:
  \begin{enumerate}
  \item[(A)] VaR$_{0.99}$, RVaR$_{0.95,0.99}$ and ES$_{0.99}$ contributions are estimated by the MC, HMC and GS methods for dimensions $d \in \{4,6,8,10\}$.
    Note that the GS is applied only to RVaR- and ES contributions, not to VaR contributions (same in the other scenarios).\\
  \item[(B)] For $d=5$, VaR$_{\alpha^{\operatorname{VaR}}}$, RVaR$_{\alpha_1^{\operatorname{RVaR}},\alpha_2^{\operatorname{RVaR}}}$ and ES$_{\alpha^{\operatorname{ES}}}$ contributions are estimated by the MC, HMC and GS methods for confidence levels $\alpha^{\operatorname{VaR}}\in \{0.9,0.99,0.999,0.9999\}$, $\alpha^{\operatorname{ES}}\in \{0.9,0.99,0.999,0.9999\}$ and $(\alpha_1^{\operatorname{RVaR}},\alpha_2^{\operatorname{RVaR}})\in \{(0.9,0.9999),(0.9,0.99),(0.99,0.999),(0.999,0.9999)\}$.\\
  \item[(C)] For $d=5$, VaR$_{0.9}$, RVaR$_{0.9,0.99}$ and ES$_{0.9}$ contributions are estimated by the MC and HMC methods with the parameters $(\epsilon_{\text{opt}},\epsilon_{\text{opt}})$ (determined by Algorithm~\ref{algorithm:adaptive:stepsize:integratio:time}) and 
  \begin{align*}
  (\epsilon,T)\in \{ (10\epsilon_{\text{opt}}, 2T_{\text{opt}}),\ (10\epsilon_{\text{opt}}, T_{\text{opt}}/2),\ (\epsilon_{\text{opt}}/10, 2T_{\text{opt}}),\ (\epsilon_{\text{opt}}/10, T_{\text{opt}}/2)\}.
  \end{align*}
  \end{enumerate}

In the MC method, the modified VaR contribution
  $\E[\bX|{\mathcal C}^{\operatorname{RVaR}}_{\alpha-\delta,\alpha+\delta}]$ with
  $\delta = 0.01$ is computed.  Moreover, if the size of the conditional sample
  for estimating RVaR and ES contributions is less than 100, then the lower
  confidence level of the crisis event is subtracted by 0.01 so that at least
  100 MC presamples are guaranteed. For the sample paths of the MCMC methods,
  ACR, ACP, and Hamiltonian errors for the HMC methods were inspected and the
  convergences of the chains were checked as in
  Section~\ref{subsec:simulation:study} and~\ref{subsec:empirical:study}.

\begin{figure}[htbp]
  \centering
  \includegraphics[width=\textwidth]{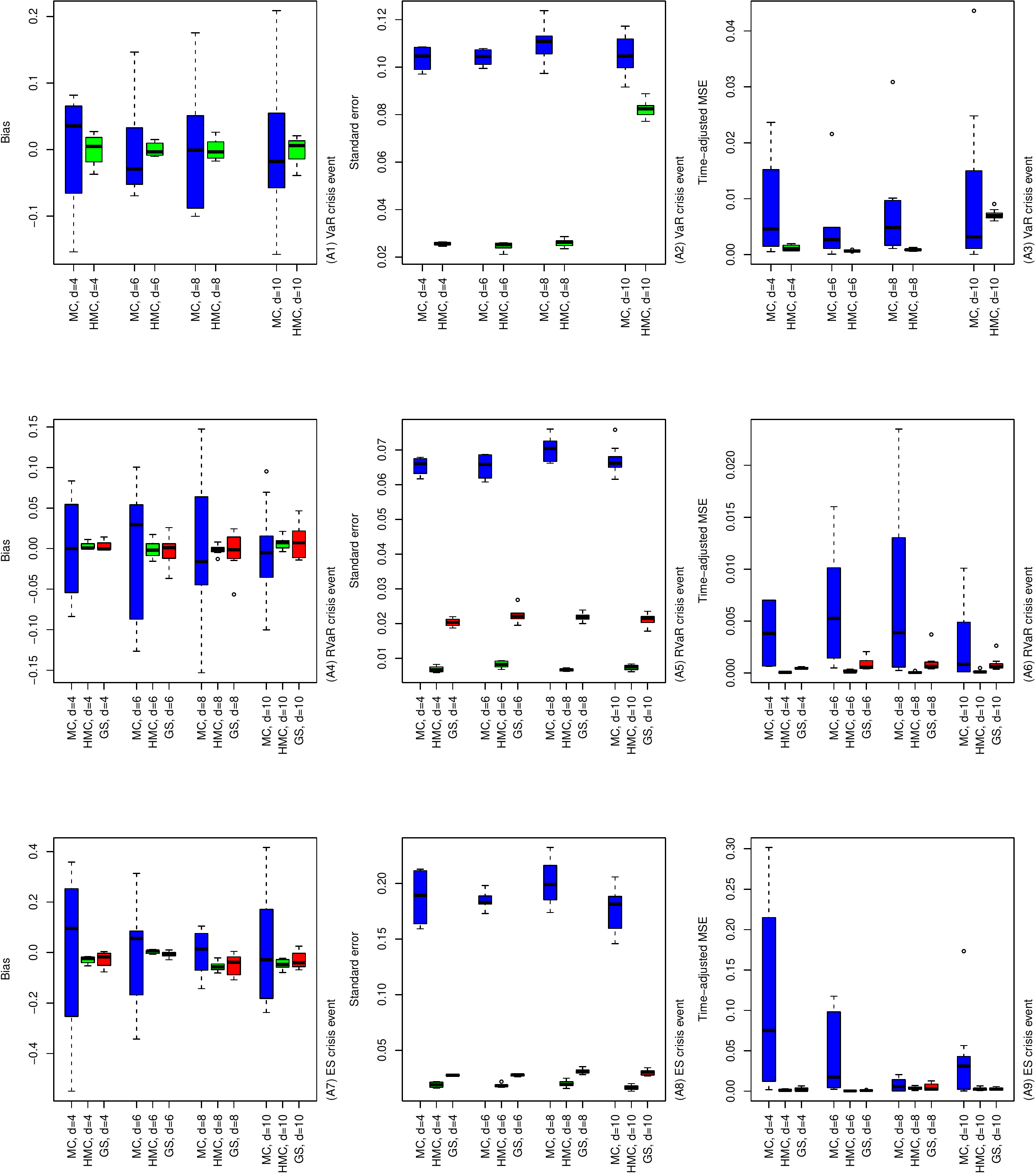}
  \caption{Bias (left), standard error (middle) and time-adjusted mean squared error (right) of
    the MC, HMC and GS estimators of risk contribution type systemic risk
    allocations under VaR$_{0.99}$ (top), RVaR$_{0.95,0.99}$ (middle) and ES$_{0.99}$ (bottom)
    crisis events.  The underlying loss distribution is $t_{\nu}(\bmu, P)$
    where $\nu=6$, $\bmu=\bzero$ and
    $P=1/12\cdot \bone_d\bone_d\T + \diag_d(11/12)$ for portfolio dimensions
    $d \in \{4,6,8,10\}$. Note that the GS method is applied only to RVaR and
    ES contributions.}
  \label{fig:comparison:dimension}
\end{figure}

\begin{figure}[htbp]
  \centering
  \includegraphics[width=\textwidth]{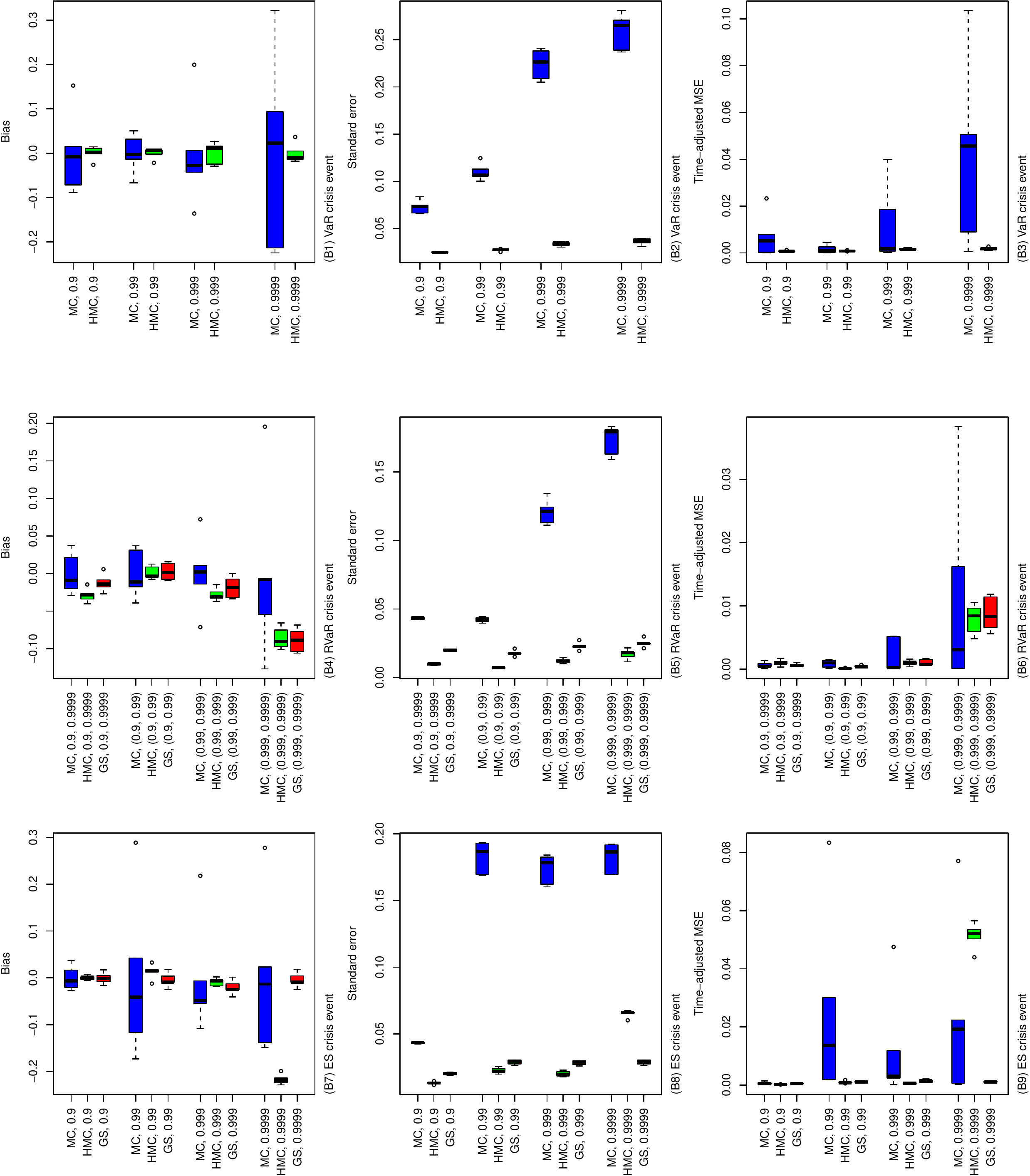}
  \caption{Bias (left), standard error (middle) and time-adjusted mean squared
    error (right) of the MC, HMC and GS estimators of risk contribution type
    systemic risk allocations with the underlying loss distribution
    $t_{\nu}(\bmu, P)$ where $\nu=6$, $\bmu=\bzero$,
    $P=1/12 \cdot\bone_d\bone_d\T + \diag_d(11/12)$ and $d=5$. The crisis
    event is taken differently as VaR$_{\alpha^{\operatorname{VaR}}}$ (top),
    RVaR$_{\alpha_1^{\operatorname{RVaR}},\alpha_2^{\operatorname{RVaR}}}$ (middle) and
    ES$_{\alpha^{\operatorname{ES}}}$ (bottom) for confidence levels
    $\alpha^{\operatorname{VaR}}\in \{0.9,0.99,0.999,0.9999\}$,
    $(\alpha_1^{\operatorname{RVaR}},\alpha_2^{\operatorname{RVaR}})\in
    \{(0.9,0.9999),(0.9,0.99),(0.99,0.999),(0.999,0.9999)\}$ and
    $\alpha^{\operatorname{ES}}\in \{0.9,0.99,0.999,0.9999\}$. Note that the GS method is
    applied only to RVaR and ES contributions.}
  \label{fig:comparison:confidence:level}
\end{figure}

\begin{figure}[htbp]
  \centering
  \includegraphics[width=\textwidth]{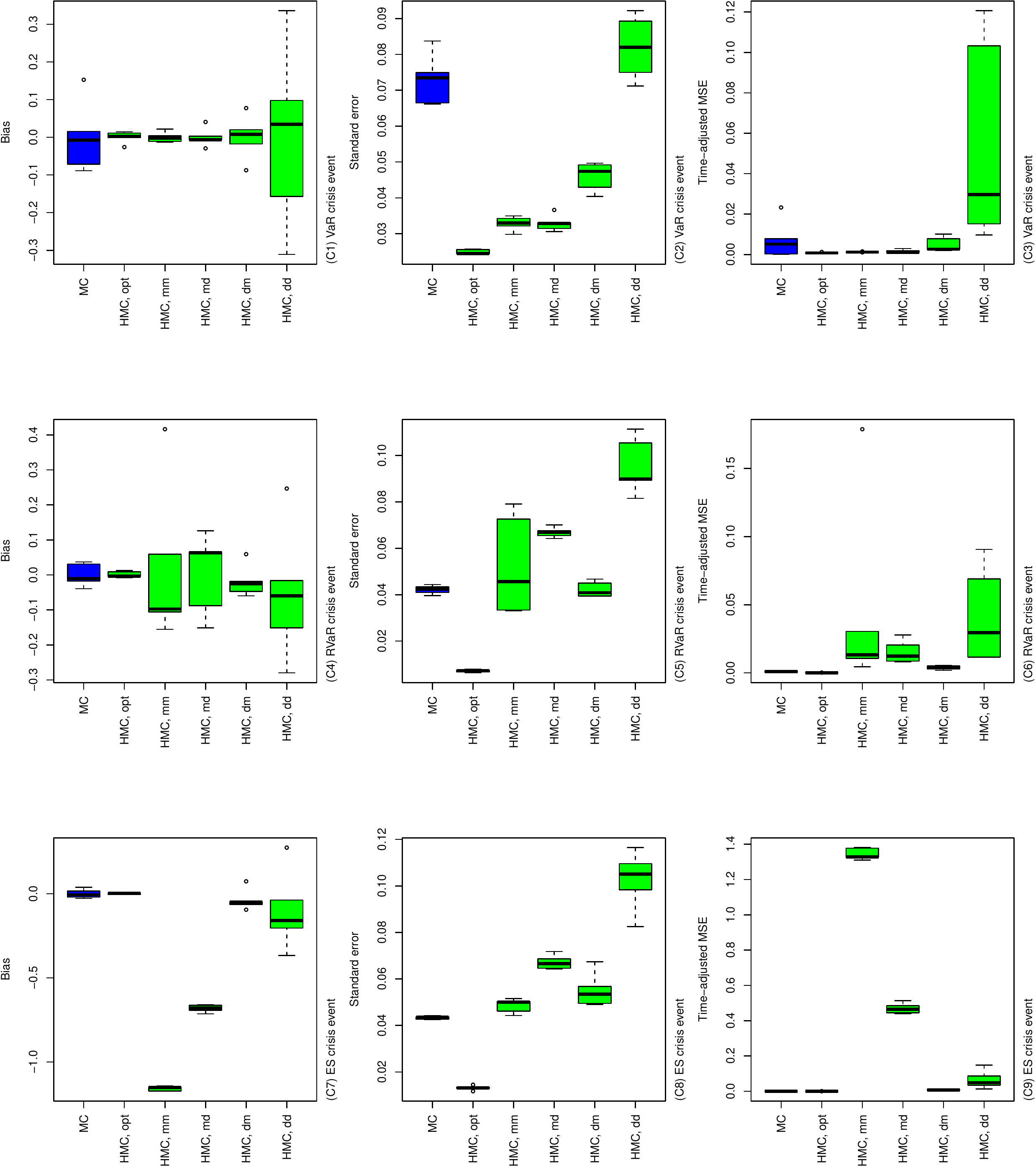}
  \caption{Bias (left), standard error (middle) and time-adjusted mean squared error (right) of
    the MC and HMC estimators of risk contribution type systemic risk
    allocations under VaR$_{0.9}$, RVaR$_{0.9,0.99}$ and ES$_{0.9}$ crisis
    events.  The underlying loss distribution is $t_{\nu}(\bmu, P)$ where
    $\nu=6$, $\bmu=\bzero$, $P=1/12 \cdot\bone_d\bone_d\T + \diag_d(11/12)$
    and $d=5$.  The parameters of the HMC method are taken as
    $(\epsilon_{\text{opt}},\epsilon_{\text{opt}})$ determined by
    Algorithm~\ref{algorithm:adaptive:stepsize:integratio:time} and
    $(\epsilon,T)\in
    \{(10\epsilon_{\text{opt}}, 2T_{\text{opt}}),\ (10\epsilon_{\text{opt}}, T_{\text{opt}}/2),\ (\epsilon_{\text{opt}}/10, 2T_{\text{opt}}),\ (\epsilon_{\text{opt}}/10, T_{\text{opt}}/2)\}$.
    In the labels of the x-axes, each of the five cases
    $(\epsilon_{\text{opt}},\epsilon_{\text{opt}}),\
    (10\epsilon_{\text{opt}},2T_{\text{opt}}),\
    (10\epsilon_{\text{opt}},T_{\text{opt}}/2),\
    (\epsilon_{\text{opt}}/10,2T_{\text{opt}})$ and
    $(\epsilon_{\text{opt}}/10,T_{\text{opt}}/2)$ is denoted by HMC.opt,
    HMC.mm, HMC.md, HMC.dm and HMC.dd, respectively.
  }\label{fig:comparison:parameters:HMC}
\end{figure}

 The results of the comparisons of (A), (B) and (C) are summarized in
  Figure~\ref{fig:comparison:dimension},~\ref{fig:comparison:confidence:level}
  and~\ref{fig:comparison:parameters:HMC}.  In
  Figure~\ref{fig:comparison:dimension}, the performance of the MC, HMC and GS
  estimators is roughly similar across dimensions from 4 to 10.  For all crisis
  events, the HMC and GS estimators outperform MC in terms of bias, standard
  error and time-adjusted MSE.  From (A5) and (A8), standard errors of the GS
  estimators are slightly higher than those of the HMC ones, which result in
  slightly improved performance of the HMC estimator over the GS in terms of
  MSE.  In Figure~\ref{fig:comparison:confidence:level}, bias, standard error
  and MSE of the MC estimator tend to increase as the probability of the
  conditioning set decreases.  This is simply because the size of the
  conditional samples in the MC method decreases proportional to the probability
  of the crisis event.  On the other hand, the HMC and GS estimators provide a
  stably better performance than MC since such sample size reduction does not
  occur.  As seen in (B4) to (B9) in the cases of RVaR$_{0.999,0.9999}$ and
  ES$_{0.9999}$, however, if the probability of the conditioning event is too
  small and/or the distribution of the MC presample is too different from the
  original conditional distribution of interest, then the parameters of the HMC
  method determined by
  Algorithm~\ref{algorithm:adaptive:stepsize:integratio:time} can be entirely
  different from optimal, which leads to a poor performance of the HMC method as
  we will see in the next scenario (C).  In
  Figure~\ref{fig:comparison:parameters:HMC}, the HMC method with optimally
  determined parameters from
  Algorithm~\ref{algorithm:adaptive:stepsize:integratio:time} is compared to
  non-optimal parameter choices.  First, the optimal HMC estimator outperforms
  MC in terms of bias, standard error and time-adjusted MSE.  On the other hand,
  from the plots in Figure~\ref{fig:comparison:parameters:HMC} we see that some of the
  non-optimal HMC estimators are significantly worse than MC.
  Therefore, a careful choice of the parameters of the HMC method is required to
  obtain an improved performance of the HMC method compared to MC.

\section{Conclusion, Limitations and Future Work}\label{sec:conclusion}
Efficient calculation of systemic risk allocations is a challenging task,
especially when the crisis event has a small probability.  To solve this problem
for models where a joint loss density is available, we proposed MCMC estimators
where a Markov chain is constructed with the conditional loss distribution given
the crisis event as the target distribution.  By using HMC and GS, efficient
simulation methods from the constrained target distribution are obtained and the
resulting MCMC estimator is expected to have a smaller standard error compared
to that of the MC estimator.  Sample efficiency is significantly improved
since the MCMC estimator is computed from samples generated directly from the
conditional distribution of interest.  Another advantage of the MCMC
  method is that its performance is less sensitive to the probability of the
  crisis event, and thus to the confidence levels of the underlying risk
  measures.  We also proposed a heuristic for determining the parameters of the
  HMC method based on the MC presamples.  Numerical experiments demonstrated
  that our MCMC estimators are more efficient than MC in terms of bias, standard
  error and time-adjusted MSE.  Stability of the MCMC estimation with respect to
  the probability of the crisis event and efficiency of the optimal parameter
  choice of the HMC method are also investigated in the experiments.  

  Based on the results in this paper, our MCMC estimators can be recommended when
  the probability of the crisis event is too small for MC to simulate
  sufficiently many samples for a statistical analysis and/or when unbiased systemic
  risk allocations under the VaR crisis event are required.  The MCMC methods
  are likely to perform well when the dimension of the portfolio is less than or
  around 10, losses are bounded from the left and the crisis event is of VaR or RVaR
  type; otherwise heavy-tailedness and computational time can become challenging.
  First, a theoretical convergence result of the HMC method is typically not
  available when the target distribution is unbounded and heavy-tailed, which is
  the case when the losses are unbounded and/or the crisis event is of ES type;
  see the case of the ES crisis event in the empirical study in
  Section~\ref{subsec:empirical:study}.  Second, both of the HMC and GS methods
  suffer from high-dimensional target distributions since the algorithms contain
  parts of steps where the computational cost linearly increases in the
  dimension. We observed that, in this case, although the MCMC estimator
  typically improves bias and standard error compared to MC, the improvement
  vanishes in terms of time-adjusted MSE due to the long computational time of
  the MCMC method. Finally, multimodality of joint loss distributions and/or the
  target distribution is also an undesirable feature since full conditional
  distributions and their inverses (which are required to implement the GS) are
  typically unavailable in the former case, and the latter case prevents the HMC method
  from efficiently exploring the entire support of the target distribution.
  Potential remedies for heavy-tailed and/or high-dimensional target
  distributions are the HMC method with a non-normal kinetic energy distribution
  and roll-back HMC; see Appendix~\ref{appendix:sec:other:mcmc:methods} for
  details.  Further investigation of HMC methods and faster methods for
  determining the HMC parameters are left for future work.

\section*{Funding}
This research was funded by NSERC through Discovery Grant RGPIN-5010-2015.



\appendix

  \section{Hamiltonian Dynamics with Boundary Reflection}\label{appendix:reflection:technique:HMC}
  In this appendix, we describe details of the HMC method
  with boundary reflection as mentioned in Section~\ref{subsubsec:HMC:reflection}.
  Let $(\bh,v)$ be the hyperplane which the trajectory of the Hamiltonian
  dynamics hit at $(\bx(t),\bp(t))$.  At this time, $(\bx(t),\bp(t))$ is
  immediately replaced by $(\bx(t),\bp_r(t))$ where $\bp_r(t)$ is the
  \emph{reflected momentum} defined by
  \begin{align*}
    \bp_r(t) = \bp_{\|}(t) - \bp_{\perp}(t),
  \end{align*}
  where $ \bp_{\|}(t)$ and $\bp_{\perp}(t)$ are such that
  $\bp(t) = \bp_{\|}(t) + \bp_{\perp}(t)$ and $ \bp_{\|}(t)$ and
  $\bp_{\perp}(t)$ are parallel and perpendicular to the hyperplane $(\bh,v)$,
  respectively.  \citet{afshar2015reflection} and
  \citet{chevallier2018hamiltonian} showed that the map
  $(\bx(t),\bp(t))\mapsto (\bx(t),\bp_r(t))$ preserves the volume and the
  Hamiltonian, and that this modified HMC method has the stationary distribution
  $\pi$.  As long as the initial position $\bx^{(0)}$ belongs to $\mathcal C$,
  the trajectory of the HMC method never violates the constraint $\mathcal C$.
  The algorithm of this HMC method with reflection is obtained by replacing the
  \textbf{Leapfrog} function call in Step 3) of Algorithm~\ref{algorithm:HMC}
  by Algorithm~\ref{algorithm:LF:reflection}.  Accordingly, the parameters of
  the hyperplanes need to be passed as input to Algorithm~\ref{algorithm:HMC}.
  \begin{algorithm}[t]
    \caption{Leapfrog method with boundary reflection}
    \label{algorithm:LF:reflection}
    \begin{algorithmic}[0]
      \INPUT  Current state $(\bx(0),\bp(0))$, stepsize $\epsilon>0$, gradients $\nabla U$ and $\nabla K$, and constraints $(\bh_m,v_m)$, $m=1,\dots,M$.\\
      \OUTPUT Updated state $(\bx(\epsilon),\bp(\epsilon))$ .\\ \vspace{1mm}
      \algrule
      \STATE 1) Update $\bp(\epsilon/2) = \bp(0) + \epsilon/2 \nabla U(\bx(0))$.\\
      \STATE 2) Set $(\bx,\bp)=(\bx(0),\bp(\epsilon/2))$, $\epsilon_\text{temp} = \epsilon$.\\
      \STATE 3) \textbf{while} $\epsilon_\text{temp}>0$\\
      \STATE \hspace{8mm} 3-1) Compute
      \begin{align*}
        \bx^{\ast}& =\bx + \epsilon_\text{temp} \nabla K(\bp),\\
        t_m &= (v_m - \bh_m\T\bx)/(\epsilon \bh_m\T\bp),\quad m=1,\dots,M.
      \end{align*}
      \STATE \hspace{8mm} 3-2) \textbf{if} $t_m \in [0,1]$ for any $m=1,\dots,M$,\\
      \STATE \hspace{8mm} \hspace{8mm} 3-2-1) Set
      \begin{align*}
        m^{\ast} &= \text{argmin}\{t_m \ |\   0\leq t_m \leq 1,\ m = 1,\dots, M\},\\
        \bx_r^{\ast}&=\bx^{\ast} -2 \frac{\bh_{m^{\ast}}\T\bx^{\ast}- v_{m^{\ast}}}{\bh_{m^{\ast}}\T\bh_{m^{\ast}}}h_{m^{\ast}},\\
        \bp_r &= \frac{\bx^{\ast}-\bx- t_{m^{\ast}}\epsilon \bp}{\epsilon(1-t_{m^{\ast}})}.
      \end{align*}
      \STATE  \hspace{8mm}  \hspace{8mm} 3-2-2) Set       $(\bx,\bp) = (\bx_{r}^{\ast},\bp_r)\quad \text{and}\quad  \epsilon_\text{temp}=(1-t_{m^{\ast}})\epsilon_\text{temp}$.\\
      \STATE  \hspace{8mm} \hspace{6mm} \textbf{else}\\
      \STATE  \hspace{8mm}  \hspace{8mm} 3-2-3) Set $(\bx,\bp)=(\bx^{\ast},\bp)$ and $\epsilon_\text{temp}=0$.\\
      \STATE \hspace{8mm} \hspace{6mm} \textbf{end if} \\
      \STATE \hspace{3mm} \textbf{end while} \\
      \STATE 4) Set $\bx(\epsilon)=\bx$ and $\bp(\epsilon) = \bp + \frac{\epsilon}{2} \nabla U(\bx)$.\\
    \end{algorithmic}
  \end{algorithm}
  In Step 3-1) of Algorithm~\ref{algorithm:LF:reflection} the time $t_m$ at
  which the trajectory hits the boundary $(\bh_m,v_m)$ is computed.  If
  $0<t_m<1$ for some $m \in \{1,\dots,M\}$, then the chain hits the boundary
  during the dynamics with length $\epsilon$.  At the smallest time
  $t_{m^{\ast}}$ among such hitting times, the chain reflects from
  $(\bx^{\ast},\bp)$ to $(\bx_{r}^{\ast},\bp_r)$ against the corresponding
  boundary $(\bh_{m^{\ast}},v_{m^{\ast}})$ as described in Step 3-2-1) of
  Algorithm~\ref{algorithm:LF:reflection}.  The remaining length of the dynamics
  is $(1-t_{m^{\ast}})\epsilon_\text{temp}$ and Step 3) is repeated until the
  remaining length becomes zero.  Other techniques of reflecting the dynamics
  are discussed in Appendix~\ref{appendix:subsec:rollback:HMC}.  

\section{Other MCMC Methods}\label{appendix:sec:other:mcmc:methods} In
  this appendix, we introduce some advanced MCMC techniques potentially
  applicable to the problem of estimating systemic risk allocations.

\subsection{Roll-Back HMC}\label{appendix:subsec:rollback:HMC}
  \citet{yi2017roll} proposed \emph{roll-back HMC (RBHMC)}, in which the
  indicator function $\bone_{[\bx \in \mathcal C]}$ in the target
  distribution~\eqref{eq:target:distribution:conditional} is replaced by a
  smooth sigmoid function so that the Hamilotonian dynamics naturally move back
  inwards when the trajectory violates the constraints.  HMC with reflection
  presented in Section~\ref{subsubsec:HMC:reflection} requires to check $M$
  boundary conditions at every iteration of the Hamiltonian dynamics.  In our
  problem the number $M$ linearly increases with the dimension $d$ in the case 
  of pure losses, which leads to a linear increase in the computational cost.
  The RBHMC method avoids such explicit boundary checks, and thus can reduce the
  computational cost of the HMC method with constrained target distributions.  Despite
  saving computational time, we observed that the RBHMC method requires a
  careful choice of the stepsize $\epsilon>0$ and the smoothness parameter of
  the sigmoid function involved, and we could not find any guidance on how to choose them to guarantee a stable performance.

\subsection{Riemannian Manifold HMC}\label{appendix:subsec:riemannian:manifold:HMC}
  \citet{livingstone2019kinetic} indicated that non-normal kinetic energy
  distributions can potentially deal with heavy-tailed target distributions.  In
  fact, the kinetic energy distribution $F_K$ can even be dependent on the
  position variable $\bx$.  For example, when
  $F_K(\cdot|\bx) = \N(\bzero,G(\bx))$ for a positive definite matrix
  $G(\bx)>0$ and $\bx \in E$, the resulting HMC method is known as \emph{Riemannian
    manifold HMC (RMHMC)} since this case is equivalent to applying HMC on the
  Riemannian manifold with metric $G(\bx)$; see \citet{girolami2011riemann}.
  Difficulties in implementing RMHMC are in the choice of metric $G$ and in the
  simulation of the Hamiltonian dynamics.  Due to the complexity of the
  Hamiltonian dynamics, simple discretization schemes such as the leapfrog
  method are not applicable, and the trajectory is updated implicitly by solving
  some system of equations; see \citet{girolami2011riemann}.  Various choices of
  the metric $G$ are studied in \citet{betancourt2013general},
  \citet{lan2014wormhole} and \citet{livingstone2014information} for different
  purposes.  Simulation of RMHMC is studied, for example, in
  \citet{byrne2013geodesic}.  

\subsection{Metropolized Gibbs
    Samplers}\label{appendix:subsec:metropolized:gibbs:sampler}
  \citet{muller1992alternatives} introduced the \emph{Metropolized Gibbs sampler
    (MGS)} in which the proposal density $q$ in the MH kernel is set to be
  $q=f_{\bY|v_1 \leq \bh\T \Y \leq v_2}$ where $\bY$ has the same marginal
  distributions as $\bX$ but a different copula $C^{q}$ for which $C_{j|-j}^{q}$
  and $C_{j|-j}^{q,-1}$ are available so that the GS can be applied to simulate
  this proposal.  This method can be used when the inversion method is not
  feasible since $C_{j|-j}$ or $C_{j|-j}^{-1}$ are not available.  Following the
  MH algorithm, the candidate is accepted with the acceptance
  probability~\eqref{eq:acceptance:probability}, which can be simply written as
    \begin{align*}
    \alpha(\bx,\tilde \bx)
    =
    \min\left\{\frac{
    c(\bm F(\tilde \bx))c^{q}(\bm F(\bx))
    }{
    c(\bm F(\bx))c^{q}(\bm F(\tilde \bx))
    },1
    \right\}.
  \end{align*}

As an example of the MGS, suppose $C$ is the Gumbel copula, for which
  the full conditional distributions cannot be inverted analytically.  One could
  then choose the survival Clayton copula as the proposal copula $C^{q}$ above.
  For this choice of copula, $q_{j|-j}$ is available by the inversion method as
  discussed in Section~\ref{subsubsec:GS:systemic:allocations}.  Furthermore,
  the acceptance probability is expected to be high especially on the upper tail
  part because the upper threshold copula of $C$ defined as
  $\Prob(\bU > \bv \ | \ \bU > \bu)$, $\bv \in [\bu,\bone]$, $\bu \in [0,1]^d$,
  $\bU \sim C$ is known to converge to that of a survival Clayton copula when
  $\lim u_j \rightarrow \infty$, $j=1,\dots,d$; see \citet{juri2002copula},
  \citet{juri2003tail}, \citet{charpentier2007lower} and
  \citet{larsson2011extremal}.


\bibliographystyle{apalike}
\bibliography{SystemicAllocationsMCMC.bib}



\end{document}